\documentclass[12pt,english, journal,draftclsnofoot,onecolumn]{IEEEtran}
\usepackage[T1]{fontenc}
\usepackage[latin9]{inputenc}
\usepackage{geometry}
\usepackage{color}
\usepackage{babel}
\usepackage{mathrsfs}
\usepackage{amsmath}
\usepackage{amssymb}
\usepackage{stackrel}
\usepackage{graphicx}
\usepackage{setspace}
\usepackage{esint}
\usepackage[unicode=true,
 bookmarks=true,bookmarksnumbered=true,bookmarksopen=true,bookmarksopenlevel=1,
 breaklinks=false,pdfborder={0 0 1},backref=false,colorlinks=true]
 {hyperref}

\makeatletter
\usepackage{enumitem}		

\usepackage{colortbl}
\usepackage{cite}
\usepackage{nomencl}
\usepackage{flushend}
\usepackage{stfloats}
\usepackage[margin=8pt,font=footnotesize,labelfont=md,labelsep=colon]{caption}

\@ifundefined{showcaptionsetup}{}{%
 \PassOptionsToPackage{caption=false}{subfig}}
\usepackage{subfig}
\makeatother

\begin{document}

\title{Error Probability Analysis and Power Allocation for Interference
Exploitation Over Rayleigh Fading Channels}

\author{Abdelhamid Salem,\textit{\normalsize{} Member, IEEE}{\normalsize{},}
and Christos Masouros, \textit{\normalsize{}Senior Member, IEEE},\\
\thanks{The authors are with the department of Electronic and Electrical Engineering,
University College London, London, UK, (emails: \{a.salem, c.masouros\}@ucl.ac.uk). 

Part of this paper has been submitted to WCNC 2020 \cite{WCNC19}.%
} }
\maketitle
\begin{abstract}
This paper considers the performance analysis of constructive interference
(CI) precoding technique in multi-user multiple-input multiple-output
(MU-MIMO) systems with a finite constellation phase-shift keying (PSK)
input alphabet. Firstly, analytical expressions for the moment generating
function (MGF) and the average of the received signal-to-noise-ratio
(SNR) are derived. Then, based on the derived MGF expression the average
symbol error probability (SEP) for the CI precoder with PSK signaling
is calculated. In this regard, new exact and very accurate asymptotic
approximation for the average SEP are provided. Building on the new
performance analysis, different power allocation schemes are considered
to enhance the achieved SEP. In the first scheme, power allocation
based on minimizing the sum symbol error probabilities (Min-Sum) is
studied, while in the second scheme the power allocation based on
minimizing the maximum SEP (Min-Max) is investigated. Furthermore,
new analytical expressions of the throughput and power efficiency
of the CI precoding in MU-MIMO systems are also derived. The numerical
results in this work demonstrate that, the CI precoding outperforms
the conventional interference suppression precoding techniques with
an up to 20dB gain in the transmit SNR in terms of SEP, and up to
15dB gain in the transmit SNR in terms of the throughput. In addition,
the SEP-based power allocation schemes provide additional up to 13dB
gains in the transmit SNR compared to the conventional equal power
allocation scheme. \end{abstract}


\section{Introduction}

Multi-user multiple-input multiple-output (MU-MIMO) communication
system has been recognized as a promising technique in wireless communication
networks \cite{MIMO1,MIMO2,MIMO3}. However, in practical implementations
the performance of MU-MIMO systems can be impacted by strong interferences
\cite{MIMO1,MIMO2,MIMO3}. Consequently, a large number of researches
have investigated the impact of the interference in MU-MIMO systems,
and several techniques have been introduced to mitigate the multi-user
interference in MU-MIMO channels \cite{MIMO3,Twway,interference1}.
For instance, in the applications when the channel state information
(CSI) is perfectly known at the base station (BS), dirty-paper coding
(DPC) technique has been proposed \cite{dpc1,chrisnewref1,chrisnewref2,angLi}.
In DPC technique the channel capacity is achieved by removing the
interference before the transmission. However, DPC is difficult if
not impossible to implement in practical communication networks, due
to its very high complexity \cite{dpc1,chrisnewref1,chrisnewref2,angLi}.
Therefore, low complexity linear precoding techniques, such as zero-forcing
(ZF), have received significant research interest \cite{zfVTC,RZF}.
Furthermore, precoding techniques based on optimization have also
widely studied and investigated in literature \cite{angli7,angli8,angli10,angli12}.
In this regard, several optimization-based schemes have been proposed
in different areas. For instance, signal-to-interference-plus-noise
ratio (SINR) balancing technique is one of these precoding schemes
that depends on maximizing the minimum SINR subject to different transmission
power constraints \cite{angli7,angli8}. In addition, minimizing the
transmission power precoding is another precoding scheme that aims
to minimize the transmission power subject to a minimum threshold
value of the SINR \cite{angli10,angli12}. 

However, all the above precoding/transmission schemes have ignored
the fact that the interference in communication systems can be beneficial
to the received signal, and thus the multi-user interference can be
exploited to further enhance the system performance. In light of this,
constructive interference (CI) precoding technique has been proposed
recently to improve the performance of down-link MIMO systems \cite{CI1,A,CI2,CI4,CI3}.
In contrast to the conventional interference mitigation techniques,
the main idea of the CI precoding scheme is to exploit the interference
that can be known to the BS/transmitter to enhance the received power
of the useful signals\cite{CI1,A,CI2,CI4,CI3}. That is, with the
knowledge of both the users\textquoteright{} channels and users\textquoteright{}
data symbols, the BS can classify the interference as constructive
and destructive. The constructive interference is the interference
that can push the received symbol deeper in the constructive/detection
region of the constellation point of interest. According to this methodology,
the preceder can be designed to make all the well known interferences
constructive to the desired symbol. The main idea of CI precoding
for PSK constellations is clarified in \cite[Section-III]{CI2}.

The concept of interference exploitation technique has been widely
considered in literature. This line of research was introduced in
\cite{CI1}, where the CI precoding technique has been proposed for
downlink MIMO systems. In this work it was shown that by exploiting
the interference signals, the system performance can be greatly enhanced
and the effective SINR can be improved without increasing the transmission
power at the BS. In \cite{A} the concept of CI was used for the first
time to design an optimization-based precoder in the form of pre-scaling.
In \cite{CI2} the authors proposed CI-based precoding schemes for
down-link MU-MIMO systems to minimize the transmit power for generic
PSK signals. The concept of CI was applied to massive-MIMO systems
in \cite{CI4}. Further work in \cite{CI3,CI5} implemented CI precoding
scheme in wireless power transfer scenarios for PSK messages, in order
to minimize the total transmit power. The authors in \cite{Luxm1,Luxm2}
considered general category of CI regions, named distance preserving
CI region, where several properties for this region have been provided.
Furthermore, recently closed-form expression for CI precoding technique
in MU-MIMO systems with PSK signaling has been derived in \cite{angLi}.
This closed-form expression has paved the way to develop theoretical
analysis of the CI technique. Based on this closed-form expression
for CI precoding, in our previous work in \cite{Tcompaper,rspaper}
closed-form expression of the achievable sum-rate of the CI precoding
technique in MU-MIMO systems has been derived and investigated.

Accordingly this paper is the first work characterizes the statistics
of CI precoding with $M$-PSK signals in MU-MIMO systems. Firstly,
analytical expressions of the moment-generating function (MGF) and
the average of the received SNR of the considered system are derived.
The derived MGF expression is then used to evaluate the average symbol
error probability (SEP). In this regard, exact SEP expression for
CI precoding with $M$-PSK signals is derived. For simplicity and
in order to provide more insight, very accurate asymptotic approximation
for the SEP is also presented. Based on the new SEP expressions, different
power allocation schemes to enhance the achieved SEP are considered.
In the first one, power allocation technique that based on minimizing
the sum symbol error probabilities (Min-Sum) is studied, while in
the second technique the power allocation that based on minimizing
the average SEP (Min-MAx) is investigated. Furthermore, the throughput
and power efficiency achieved by the CI precoding in MU-MIMO systems
are also studied. In this context, new analytical expressions of the
throughput and power efficiency are provided.

For clarity, we summarize the main contributions of this paper as: 
\begin{enumerate}
\item New and explicit analytical expressions for the MGF and the average
of the received SNR for CI precoding technique under $M$-PSK inputs
are derived.
\item New, exact, and explicit expression of the average SEP for CI precoding
with $M$-PSK is derived. For simplicity and mathematical tractability,
new and very accurate asymptotic approximation of the SEP is also
provided.
\item Two power allocation schemes to improve the SEP and enhance the system
performance are proposed. In the first one, we consider power allocation
technique that aims to minimize the sum symbol error probabilities
subject to total power constraint. Whilst in the second scheme, we
study power allocation technique that aims to minimize the maximum
SEP subject to total power constraint.
\item Based on the above analysis, closed form expression of the power allocation
factors are presented.
\item New and explicit analytical expressions for the throughput and power
efficiency for the CI precoding in MU-MIMO systems under $M$-PSK
inputs are also derived.
\end{enumerate}
The numerical results in this paper show that, for a given SEP the
CI precoding can provide up to 20dB gain in the transmit SNR compared
to the conventional interference suppression precoding techniques.
In addition, increasing the transmit SNR, number of users and number
of BS antennas always enhance the achieved SEP. Furthermore, by using
the derived analysis specifically tailored power allocation schemes
provide additional up to 13dB gains in the transmit SNR compared to
the conventional transmission scheme. Finally, the CI precoding outperforms
the conventional interference suppression precoding technique in terms
of throughput for a wide range with an up to 15dB gain in the transmit
SNR.

Next, Section \ref{sec:System-Model} describes the MU-MIMO system
model. Section \ref{sec:MGF-DERIVATION}, derives the analytical expressions
for the moment generating function and the average received SNR. Section
\ref{sec:Average-Symbol-Error} derives the exact and approximated
analytical expressions for the average symbol error probability. Section
\ref{sec:Error-Minimization-Through}, considers symbol error minimization
through different power allocation schemes, minimizing the sum symbol
error probabilities and minimizing the maximum symbol error probability.
Section \ref{sec:Throughput-and-Power}, considers the throughput
and power efficiency for the CI precoding in MU-MIMO systems. The
graphical illustrations of the results are presented and discussed
in Section \ref{sec:Numerical-Results}. Finally, our conclusions
are presented in Section \ref{sec:Conclusions}.

\section{System Model\label{sec:System-Model}}

We consider a down-link MU-MIMO system, consisting of $N$-antennas
BS communicating simultaneously with $K$ single antenna users. In
this model, the channels between the BS and the users are assumed
to be independent identically distributed (i.i.d) Rayleigh fading
channels. The $K\times N$ channel matrix between the BS and the $K$
users is denoted by $\mathbf{H}$, which can be expressed as $\mathbf{H}=\mathbf{D}^{1/2}\tilde{\mathbf{H}}$
where $\tilde{\mathbf{H}}$ is $K\times N$ matrix has i.i.d $\mathcal{CN}\left(\text{0,}1\right)$
elements which represent small scale fading coefficients and $\mathbf{D}$
is $K\times K$ a diagonal matrix with$\left[\mathbf{D}\right]_{kk}=\varpi_{k}=d_{k}^{-m}$
and $m$ is the path-loss exponent, which represents the path-loss
attenuation. It is also assumed that the CSI is perfectly known at
the BS. The received signal at the $k^{th}$ user in the considered
system can be written as,

\begin{equation}
y_{k}=\mathbf{h}_{k}\mathbf{W}\mathbf{x}+n_{k}\label{eq:1}
\end{equation}

\noindent where $\mathbf{x}$ is the PSK-modulated signal vector,
$\mathbf{W}$ is the precoding matrix, $\mathbf{h}_{k}$ is the channel
vector of user $k$, and $n_{k}$ is the additive wight Gaussian noise
(AWGN) at the $k^{th}$ user, $n_{k}\sim\mathcal{CN}\left(\text{0, }\sigma_{k}^{2}\right)$.
 The closed-form expression for CI precoding with PSK signaling can
be expressed as \cite{angLi,Tcompaper,rspaper}

\begin{equation}
\mathbf{W}=\frac{1}{K}\beta\,\mathbf{H}^{H}\left(\mathbf{H}\mathbf{H}^{H}\right)^{-1}\textrm{diag}\left\{ \mathbf{V}^{-1}\mathbf{u}\right\} \mathbf{x}\mathbf{x}^{H},\label{eq:2}
\end{equation}

\noindent where $\beta=\sqrt{P_{p}}\beta_{p}$, $P_{p}$ is the total
transmit power and $\beta_{P}$ is the scaling factor, $\beta_{p}=\sqrt{\frac{1}{\mathbf{u}^{H}\mathbf{V}^{-1}\mathbf{u}}}$,
while $\mathbf{V}=\textrm{diag}\left(\mathbf{x}^{H}\right)\left(\mathbf{H}\mathbf{H}^{H}\right)^{-1}\textrm{diag}\left(\mathbf{x}\right)$
and $\mathbf{1}^{H}\mathbf{u}=1$. As by CI precoding the resulting
interference contributes to the useful signal power, it has been shown
that the received SNR at user $k$ using CI precoding technique can
be written as \cite{CI2,CI5}

\begin{equation}
\gamma_{k}=\frac{\left|\boldsymbol{h}_{k}\mathbf{W}\mathbf{x}\right|^{2}}{\sigma_{k}^{2}}\label{eq:3-1}
\end{equation}

In the following sections we will study the statistics of the received
SNR and analyze the performance of CI precoding technique in details.

\section{MGF and Average SNR for CI Precoding \label{sec:MGF-DERIVATION}}

In this section, we derive the MGF and the average SNR expressions
of the considered MU-MIMO system. To start with, by substituting (\ref{eq:2})
into (\ref{eq:3-1}), the SNR at user $k$ using CI precoding technique
can be expressed as

\begin{equation}
\gamma_{k}=\frac{\left|\boldsymbol{h}_{k}\frac{\sqrt{P_{p}}\beta_{p}}{K}\,\mathbf{H}^{H}\left(\mathbf{H}\mathbf{H}^{H}\right)^{-1}\textrm{diag}\left\{ \mathbf{V}^{-1}\mathbf{u}\right\} \mathbf{x}\right|^{2}}{\sigma_{k}^{2}}\label{eq:3-1-1}
\end{equation}

For simplicity but without loss of generality, the scaling factor
$\beta_{p}$ is designed to constrain the long-term total transmit
power at the BS, and thus it can be expressed as \cite{Twway,angLi}
$\beta_{p}=\frac{1}{\sqrt{\mathcal{E}\left\{ \mathbf{u}^{H}\mathbf{V}^{-1}\mathbf{u}\right\} }}$.
Since $\left(\mathbf{H}\mathbf{H}^{H}\right)$ has Wishart distribution,
we can find that, $\beta_{p}=\frac{1}{\sqrt{\mathbf{u}^{H}\textrm{diag}\left(\mathbf{x}^{H}\right)^{-1}\, N\mathbf{\Sigma}\left(\textrm{diag}\left(\mathbf{x}\right)\right)^{-1}\mathbf{u}}}$,
where $\Sigma=\textrm{D}$ \cite{bookaspects}. The last formula
in (\ref{eq:3-1-1}) can be expressed also as

\begin{equation}
\gamma_{k}=\frac{\left|\frac{\sqrt{P_{p}}\beta_{p}}{K}\boldsymbol{b}\mathbf{A}\mathbf{u}x_{k}\right|^{2}}{\sigma_{k}^{2}}\label{eq:3}
\end{equation}

\noindent where $\mathbf{b}=\mathbf{a}_{k}$ , $\mathbf{a}_{k}$ is
a $1\times K$ vector the $k^{th}$ element of this vector is one
and all the other elements are zeros, and $\mathbf{A}=\mathbf{V}^{-1}$.
We can re-write the SNR expression in (\ref{eq:3}) as 

\begin{equation}
\gamma_{k}=\frac{\left|\frac{\sqrt{P_{p}}\beta_{p}}{K}\mathbf{b}\Sigma\mathbf{u}\frac{\boldsymbol{b}\mathbf{A}\mathbf{u}}{\mathbf{b}\Sigma\mathbf{u}}x_{k}\right|^{2}}{\sigma_{k}^{2}}=\alpha_{k}\left|g\right|^{2}\label{eq:7}
\end{equation}

\noindent where $\alpha_{k}=\frac{\left|\frac{\sqrt{P_{p}}\beta_{p}}{K}\mathbf{b}\Sigma\mathbf{u}\right|^{2}}{\sigma_{k}^{2}}$
and $g=\frac{\boldsymbol{b}\mathbf{A}\mathbf{u}}{\mathbf{b}\Sigma\mathbf{u}}$.
It was shown in literature that, the distribution of $g=\frac{\boldsymbol{b}\mathbf{A}\mathbf{u}}{\mathbf{b}\Sigma\mathbf{u}}$
can be approximated to Gamma distribution with shape parameter  $\nu$
and scale parameter $\theta$, $g\thicksim\Gamma\left(\nu,\theta\right)$
\cite{bookaspects,eaton}. Consequently, the received SNR, $\gamma_{k}$,
can be approximated to General Gamma distribution $\Gamma\left(p,d,a\right)$
with $p=\frac{1}{2}$, $d=\frac{\nu}{2}$ and $a=\theta^{2}$. Therefore,
the cumulative distribution function (CDF) and the probability density
function (PDF) of the received SNR, $\gamma_{k}$, can be written,
respectively, as

\begin{equation}
F_{\gamma_{k}}\left(\gamma\right)=\left(\frac{\varphi\left(d/p,\left(\gamma/a\right)^{p}\right)}{\Gamma\left(d/p\right)}\right)\;\textrm{and}\; f_{\gamma_{k}}\left(\gamma\right)=\left(\frac{\left(\frac{p}{a^{d}}\right)\gamma^{d-1}e^{-\left(\frac{\gamma}{a}\right)^{p}}}{\Gamma\left(\frac{d}{p}\right)}\right)\label{eq:9}
\end{equation}

\noindent where $\varphi\left(.\right)$ is the lower incomplete Gamma
function. Now the MGF of the received SNR, $\gamma_{k}$, can be derived
as

\begin{equation}
\mathcal{M}_{\gamma_{k}}\left(z\right)=\stackrel[0]{\infty}{\int}e^{-z\gamma}f_{\gamma_{k}}\left(\gamma\right)d\gamma\label{eq:14}
\end{equation}

Substituting the PDF in (\ref{eq:9}) into (\ref{eq:14}), we can
find 

\begin{equation}
\mathcal{M}_{\gamma}\left(z\right)=\stackrel[0]{\infty}{\int}e^{-z\gamma}\left(\frac{\left(\frac{p}{a^{d}}\right)\gamma^{d-1}e^{-\left(\frac{\gamma}{a}\right)^{p}}}{\Gamma\left(\frac{d}{p}\right)}\right)d\gamma
\end{equation}

Applying Gaussian Quadrature rule, the MGF can be simplified to 

\begin{equation}
\mathcal{M}_{\gamma}\left(z\right)=\stackrel[i=1]{n}{\sum}\textrm{H}_{i}\left(\frac{\left(\frac{p}{a^{d}}\right)e^{-\left(z-1\right)\gamma_{i}}\left(\gamma_{i}\right)^{d-1}e^{-\left(\frac{\gamma_{i}}{a}\right)^{p}}}{\Gamma\left(\frac{d}{p}\right)}\right)\label{eq:16}
\end{equation}

\noindent where $\gamma_{i}$ and $\textrm{H}_{i}$ are the $i^{th}$
zero and the weighting factor of the Laguerre polynomials, respectively
\cite{book2}. Alternatively, using Gamma distribution we can find

\begin{equation}
\mathcal{M}_{\gamma}\left(z\right)=\stackrel[i=1]{n}{\sum}\textrm{H}_{i}e^{-zP_{p}\zeta_{k}\left|g_{i}\right|^{2}}\left(\frac{g_{i}^{N-1}}{\left(N-1\right)!}\right)\label{eq:14-1}
\end{equation}

\noindent where $\zeta_{k}=\frac{\left|\frac{\beta_{p}}{K}\mathbf{b}\Sigma\mathbf{u}\right|^{2}}{\sigma_{k}^{2}}$,
$g_{i}$ here is the $i^{th}$ zero of the Laguerre polynomials \cite{book2}.

\subsection{\noindent Average SNR\label{sub:Average-SINR}}

The average SNR of CI precoder can be obtained from the first derivative
of $\mathcal{M}_{\gamma}\left(z\right)$ expressions evaluated at
$z=0$. Hence, the average SNR can be calculated by

\begin{equation}
\bar{\gamma}_{k}=\left.\frac{\partial\mathcal{M}_{\gamma}\left(z\right)}{\partial z}\right|_{z=0}
\end{equation}

\begin{equation}
\bar{\gamma}_{k}=\stackrel[i=1]{n}{\sum}\textrm{H}_{i}\frac{\partial}{\partial z}\left.\left(\frac{\left(\frac{p}{a^{d}}\right)e^{-\left(z-1\right)\gamma_{i}}\left(\gamma_{i}\right)^{d-1}e^{-\left(\frac{\gamma_{i}}{a}\right)^{p}}}{\Gamma\left(\frac{d}{p}\right)}\right)\right|_{z=0}
\end{equation}

Using a standard approach, the average of the SNR can be expressed
as

\begin{equation}
\bar{\gamma}_{k}=\stackrel[0]{\infty}{\int}\gamma\, f_{\gamma_{k}}\left(\gamma\right)d\gamma\label{eq:10}
\end{equation}

Substituting the PDF in (\ref{eq:9}) into (\ref{eq:10}) we can get,

\begin{equation}
\bar{\gamma}_{k}=\left(\frac{a\Gamma\left(\frac{1+d}{p}\right)}{\Gamma\left(\frac{d}{p}\right)}\right)=\frac{\alpha_{k}^{2}\Gamma\left(N+2\right)}{\Gamma\left(N\right)}
\end{equation}

\section{Average Symbol Error Probability (SEP)\label{sec:Average-Symbol-Error}}

In this section we calculate the average SEP for CI precoding with
$M$-PSK signaling using a standard approach provided in literature
\cite[(5.67)]{salimbook,Mckaypaper}. The average SEP of $M$-PSK
can be calculated by \cite[(5.67)]{salimbook}

\begin{equation}
P_{e,k}=\frac{1}{\pi}\stackrel[0]{\frac{\pi\left(M-1\right)}{M}}{\int}\mathcal{M}_{\gamma}\left(-\frac{\sin^{2}\left(\frac{\pi}{M}\right)}{\sin^{2}\Phi}\right)d\Phi\label{eq:19-1}
\end{equation}

\noindent  Next we will provide exact and approximated formulas
to calculate the average SEP for MU-MIMO transmission using CI precoding
technique.

\subsection{Exact SEP}

For simplicity (\ref{eq:19-1}) can be written as 

\begin{equation}
P_{e,k}=\frac{1}{\pi}\stackrel[0]{\Theta}{\int}\mathcal{M}_{\gamma}\left(z\right)d\theta\label{eq:20-1}
\end{equation}

\noindent where $\Theta=\frac{\pi\left(M-1\right)}{M}$ and $z=-\frac{\sin^{2}\left(\frac{\pi}{M}\right)}{\sin^{2}\Phi}$.
By Substituting (\ref{eq:16}) into (\ref{eq:20-1}), we can get 

\begin{equation}
P_{e,k}=\frac{1}{\pi}\stackrel[i=1]{n}{\sum}\stackrel[0]{\Theta}{\int}\frac{\textrm{H}_{i}}{zP_{p}\zeta_{k}}\left(\frac{\left(\frac{p}{a^{d}}\right)\left(\frac{\upsilon_{i}}{zP_{p}\zeta_{k}}\right)^{d-1}e^{-\left(\frac{\upsilon_{i}}{azP_{p}\zeta_{k}}\right)^{p}}}{\Gamma\left(\frac{d}{p}\right)}\right)d\Phi
\end{equation}

and

\begin{equation}
P_{e,k}=\frac{1}{\pi}\stackrel[i=1]{n}{\sum}\stackrel[0]{\Theta}{\int}\textrm{H}_{i}\frac{\left(\frac{p}{a^{d}}\right)\left(\upsilon_{i}\right)^{d-1}e^{-\left(\frac{\upsilon_{i}}{azP_{p}\zeta_{k}}\right)^{p}}}{\left(zP_{p}\zeta_{k}\right)^{d}\Gamma\left(\frac{d}{p}\right)}d\Phi\label{eq:25}
\end{equation}

Using Gamma distribution we can also find

\begin{equation}
P_{e,k}=\frac{1}{\pi}\stackrel[i=1]{n}{\sum}\stackrel[0]{\Theta}{\int}\textrm{H}_{i}e^{-zP_{p}\zeta_{k}\left|g_{i}\right|^{2}}\left(\frac{g_{i}^{N-1}}{\left(N-1\right)!}\right)d\Phi
\end{equation}

As we can notice from the derived SEP equations, the derived SEP expressions
are represented only with single integration which can be approximated
efficiently using numerical integration methods. In order to provide
more insights, in the next sub-section we derive an approximation
of the average SEP, which is shown in the numerical results to be
very accurate.

\subsection{Approximate SEP\label{sub:Approximate-SEP}}

Here we derive an approximation expression of the average SEP of the
considered scenario. Firstly, (\ref{eq:20-1}) can be written as

\begin{equation}
P_{e,k}=\mathcal{E}\left[\frac{1}{\pi}\stackrel[0]{\frac{\pi}{2}}{\int}\exp\left(-\frac{\sin^{2}\left(\frac{\pi}{M}\right)}{\sin^{2}\theta}\right)d\theta\right.\left.+\frac{1}{\pi}\stackrel[\frac{\pi}{2}]{\Theta}{\int}\exp\left(-\frac{\sin^{2}\left(\frac{\pi}{M}\right)}{\sin^{2}\theta}\right)d\theta\right]\label{eq:18}
\end{equation}

Now, the first term in (\ref{eq:18}) can be approximated by \cite{Mckaypaper,approximationSEP}

\begin{equation}
\frac{1}{\pi}\stackrel[0]{\frac{\pi}{2}}{\int}\exp\left(-\frac{\sin^{2}\left(\frac{\pi}{M}\right)}{\sin^{2}\theta}\right)d\theta\approx\frac{1}{12}e^{\left(-\sin^{2}\left(\frac{\pi}{M}\right)\right)}+\frac{1}{4}e^{\left(-\frac{4\sin^{2}\left(\frac{\pi}{M}\right)}{3}\right)}\label{eq:19}
\end{equation}

Similarly, the second term in (\ref{eq:18}) can be approximated as
\cite{Mckaypaper,approximationSEP} 
\begin{equation}
\frac{1}{\pi}\stackrel[\frac{\pi}{2}]{\Theta}{\int}\exp\left(-\frac{\sin^{2}\left(\frac{\pi}{M}\right)}{\sin^{2}\theta}\right)d\theta\approx\frac{1}{2\pi}\left(e^{\left(-\sin^{2}\left(\frac{\pi}{M}\right)\right)}+\frac{1}{4}e^{\left(-\frac{\sin^{2}\left(\frac{\pi}{M}\right)}{\sin^{2}\Theta}\right)}\right)\left(\Theta-\frac{\pi}{2}\right)\label{eq:20}
\end{equation}

Now substituting (\ref{eq:19}) and (\ref{eq:20}) into (\ref{eq:18}),
we can obtain approximated expression of SEP as \cite{Mckaypaper,approximationSEP}

\begin{equation}
P_{e,k}=\mathcal{E}\left[\frac{1}{12}e^{\left(-\sin^{2}\left(\frac{\pi}{M}\right)\right)}+\frac{1}{4}e^{\left(-\frac{4\sin^{2}\left(\frac{\pi}{M}\right)}{3}\right)}+\frac{1}{2\pi}\left(e^{\left(-\sin^{2}\left(\frac{\pi}{M}\right)\right)}+\frac{1}{4}e^{\left(-\frac{\sin^{2}\left(\frac{\pi}{M}\right)}{\sin^{2}\Theta}\right)}\right)\left(\Theta-\frac{\pi}{2}\right)\right]
\end{equation}

which can be written as

\[
P_{e,k}=\frac{1}{12}\mathcal{M}_{\gamma}\left(\sin^{2}\left(\frac{\pi}{M}\right)\right)+\frac{1}{4}\mathcal{M}_{\gamma}\left(\frac{4\sin^{2}\left(\frac{\pi}{M}\right)}{3}\right)
\]

\begin{equation}
+\frac{1}{2\pi}\left(\mathcal{M}_{\gamma}\left(\sin^{2}\left(\frac{\pi}{M}\right)\right)+\frac{1}{4}\mathcal{M}_{\gamma}\left(\frac{\sin^{2}\left(\frac{\pi}{M}\right)}{\sin^{2}\Theta}\right)\right)\left(\Theta-\frac{\pi}{2}\right)
\end{equation}

and

\begin{equation}
P_{e,k}=\left(\frac{\Theta}{2\pi}-\frac{1}{6}\right)\mathcal{M}_{\gamma}\left(\sin^{2}\left(\frac{\pi}{M}\right)\right)+\frac{1}{4}\mathcal{M}_{\gamma}\left(\frac{4\sin^{2}\left(\frac{\pi}{M}\right)}{3}\right)+\left(\frac{\Theta}{2\pi}-\frac{1}{4}\right)\mathcal{M}_{\gamma}\left(\frac{\sin^{2}\left(\frac{\pi}{M}\right)}{\sin^{2}\Theta}\right)\label{eq:29}
\end{equation}

Finally using the derived formula in (\ref{eq:29}), the approximated
expression of the average SEP for MU-MIMO system using CI precoding
technique can be written as,

\[
P_{e,k}=\left(\frac{\Theta}{2\pi}-\frac{1}{6}\right)\stackrel[i=1]{n}{\sum}\frac{\textrm{H}_{i}}{\sin^{2}\left(\frac{\pi}{M}\right)P_{p}\zeta_{k}}\left(\frac{\left(\frac{p}{a^{d}}\right)\left(\frac{\gamma_{i}}{\sin^{2}\left(\frac{\pi}{M}\right)P_{p}\zeta_{k}}\right)^{d-1}e^{-\left(\frac{\gamma_{i}}{azP_{p}\zeta_{k}}\right)^{p}}}{\Gamma\left(\frac{d}{p}\right)}\right)
\]

\[
+\frac{1}{4}\stackrel[i=1]{n}{\sum}\frac{3\textrm{H}_{i}}{4\sin^{2}\left(\frac{\pi}{M}\right)P_{p}\zeta_{k}}\left(\frac{\left(\frac{p}{a^{d}}\right)\left(\frac{3\gamma_{i}}{4\sin^{2}\left(\frac{\pi}{M}\right)P_{p}\zeta_{k}}\right)^{d-1}e^{-\left(\frac{\gamma_{i}}{azP_{p}\zeta_{k}}\right)^{p}}}{\Gamma\left(\frac{d}{p}\right)}\right)
\]

\begin{equation}
+\left(\frac{\Theta}{2\pi}-\frac{1}{4}\right)\stackrel[i=1]{n}{\sum}\frac{\textrm{H}_{i}\sin^{2}\Theta}{\sin^{2}\left(\frac{\pi}{M}\right)P_{p}\zeta_{k}}\left(\frac{\left(\frac{p}{a^{d}}\right)\left(\frac{\gamma_{i}\sin^{2}\Theta}{\sin^{2}\left(\frac{\pi}{M}\right)P_{p}\zeta_{k}}\right)^{d-1}e^{-\left(\frac{\gamma_{i}}{azP_{p}\zeta_{k}}\right)^{p}}}{\Gamma\left(\frac{d}{p}\right)}\right)\label{eq:29-1}
\end{equation}

The numerical results show that the approximation expression in (\ref{eq:29-1})
is very tight to the exact one.

\section{Error Minimization Through Power Allocation \label{sec:Error-Minimization-Through}}

Equal power allocation (EPA) is not an optimal scheme for allocating
the total transmission power between the users in communication systems,
particularly when there is a notable disparity of channel strengths
among the users. Therefore, the main aim of this section is to employ
the above analytical results to improve the performance of the CI
precoding technique with non-equal power allocation, under the assumption
of total power constraint. The considered approaches here seeking
to explain the potential gain attained in the average SEP performance
if the total available power is allocated more efficiently compared
to the baseline EPA scheme. Firstly, we study power allocation scheme
based on minimizing the sum symbol error probabilities, Min-Sum. In
the second scheme we consider the power allocation based on minimizing
the maximum SEP, Min-Max.

\subsection{Min-Sum SEP}

As we can see from the previous sections the derived SEP expressions
are functions of the power allocated at the BS and thus this amount
of power can be allocated in order to enhance the quality of the BS
transmission. Here we consider the power allocation strategy that
minimizes the total SEP of the considered system subject to the sum-power
constraint. Accordingly, the corresponding optimization problem can
be formulated as 

\[
\underset{\boldsymbol{a}}{\min}\quad\boldsymbol{1}_{K}^{T}\,\boldsymbol{\mathfrak{p}}
\]

\begin{equation}
S.t:\stackrel[k=1]{K}{\sum}a_{k}=1,\; a_{k}\geq0\label{eq:33}
\end{equation}

\noindent where $\boldsymbol{\mathfrak{p}}=\left[P_{e,1},...,P_{e,k},....,P_{e,K}\right]^{T}$
is the users SEP vector and $\boldsymbol{a}=\left[a_{1},....,a_{k},...,a_{K}\right]$
is the relative power allocation vector. This optimization problem
in (\ref{eq:33}) can be formulated in a simpler way as 

\[
\underset{\boldsymbol{a}}{\min}\quad\stackrel[k=1]{K}{\sum}P_{e,k}
\]

\begin{equation}
S.t:\stackrel[k=1]{K}{\sum}a_{k}=1,\; a_{k}\geq0\label{eq:35-2}
\end{equation}

For simplicity, substituting (\ref{eq:14-1}) into the derived SEP
expression in (\ref{eq:29}) and (\ref{eq:35-2}), we can get

\[
\underset{a_{k}}{\min}\quad\stackrel[k=1]{K}{\sum}\left\{ c_{1}\left[\stackrel[i=1]{n}{\sum}\vartheta_{i}e^{-z_{1}a_{k}P_{p}\zeta_{k}\left|g_{i}\right|^{2}}\right]+c_{2}\left[\stackrel[i=1]{n}{\sum}\vartheta_{i}e^{-z_{2}a_{k}P_{p}\zeta_{k}\left|g_{i}\right|^{2}}\right]+c_{3}\left[\stackrel[i=1]{n}{\sum}\vartheta_{i}e^{-z_{3}a_{k}P_{p}\zeta_{k}\left|g_{i}\right|^{2}}\right]\right\} 
\]

\begin{equation}
S.t:\stackrel[k=1]{K}{\sum}a_{k}=1,\; a_{k}\geq0\label{eq:35-1}
\end{equation}

\noindent where $c_{1}=\frac{\left(\frac{\left(M-1\right)}{2M}-\frac{1}{6}\right)}{\left(N-1\right)!}$,
$c_{2}=\frac{1}{4\left(N-1\right)!}$, $c_{3}=\frac{\left(\frac{\left(M-1\right)}{2M}-\frac{1}{4}\right)}{\left(N-1\right)!}$,
$\vartheta_{i}=g_{i}^{N-1}\textrm{H}_{i}$, $z_{1}=\sin^{2}\left(\frac{\pi}{M}\right)$,
$z_{2}=\frac{4\sin^{2}\left(\frac{\pi}{M}\right)}{3}$ and $z_{3}=\frac{\sin^{2}\left(\frac{\pi}{M}\right)}{\sin^{2}\frac{\pi\left(M-1\right)}{M}}$.
The function in (\ref{eq:35-1}) is convex in the parameters $a_{k}$
over the feasible set defined by linear power ratio constraints, $\frac{\partial^{2}}{\partial a_{k}^{2}}P_{e,k}>0$
for $a_{k}>0$. Therefore, the optimization problem (\ref{eq:35-1})
can be solved using CVX and other numerical software tools. However,
to develop some insights for the power allocation policy we can consider
numerical solution of this problem as follows. Following the definitions
in \cite{Boyd04}, the Lagrangian of this optimization problem in
(\ref{eq:35-1}) can be written as,

\begin{equation}
\mathfrak{L}\left(\boldsymbol{\mathfrak{p}},\lambda\right)=\boldsymbol{1}_{K}^{T}\,\boldsymbol{\mathfrak{p}}+\lambda\left(\stackrel[k=1]{K}{\sum}a_{k}-1\right)
\end{equation}

\noindent where $\lambda$ is the Lagrange multiplier satisfying the
power constraint. Therefore, the power allocation solution can be
found from the conditions

\begin{equation}
\frac{\partial}{\partial\lambda}\mathfrak{L}\left(\boldsymbol{\mathfrak{p}},\lambda\right)=\left(\stackrel[k=1]{K}{\sum}a_{k}-1\right)=0\label{eq:34}
\end{equation}

\begin{equation}
\frac{\partial}{\partial a_{k}}\mathfrak{L}\left(\boldsymbol{\mathfrak{p}},\lambda\right)=\lambda-\psi_{k}=0\label{eq:35}
\end{equation}

\noindent where $\psi_{k}=c_{1}\left[\stackrel[i=1]{n}{\sum}\omega_{i1,k}\vartheta_{i}e^{-\omega_{i1,k}a_{k}}\right]+c_{2}\left[\stackrel[i=1]{n}{\sum}\omega_{i2,k}\vartheta_{i}e^{-a_{k}\omega_{i2,k}}\right]+c_{3}\left[\stackrel[i=1]{n}{\sum}\omega_{i3,k}\vartheta_{i}e^{-a_{k}\omega_{i3,k}}\right]$,
$\omega_{ij,k}=z_{j}P_{p}\zeta_{k}\left|g_{i}\right|^{2},\, j=1,2,3.$
From (\ref{eq:35}), we can notice that $\psi_{k}=\psi_{k-1}=...=\psi_{1}$,
so that

\[
c_{1}\left[\stackrel[i=1]{n}{\sum}\omega_{i1,k}\vartheta_{i}e^{-\omega_{i1,k}a_{k}}\right]+c_{2}\left[\stackrel[i=1]{n}{\sum}\omega_{i2,k}\vartheta_{i}e^{-a_{k}\omega_{i2,k}}\right]+c_{3}\left[\stackrel[i=1]{n}{\sum}\omega_{i3,k}\vartheta_{i}e^{-a_{k}\omega_{i3,k}}\right]
\]

\begin{equation}
=c_{1}\left[\stackrel[i=1]{n}{\sum}\omega_{i1,1}\vartheta_{i}e^{-\omega_{i1,1}a_{1}}\right]+c_{2}\left[\stackrel[i=1]{n}{\sum}\omega_{i2,1}\vartheta_{i}e^{-a_{1}\omega_{i2,1}}\right]+c_{3}\left[\stackrel[i=1]{n}{\sum}\omega_{i3,1}\vartheta_{i}e^{-a_{1}\omega_{i3,1}}\right]
\end{equation}

\noindent Considering the first-order Laguerre polynomial, we can
get

\[
c_{1}\left[\omega_{11,k}e^{-\omega_{11,k}a_{k}}\right]+c_{2}\left[\omega_{12,k}e^{-a_{k}\omega_{12,k}}\right]+c_{3}\left[\omega_{13,k}e^{-a_{k}\omega_{13,k}}\right]
\]

\begin{equation}
=c_{1}\left[\omega_{11,1}e^{-\omega_{11,1}a_{1}}\right]+c_{2}\left[\omega_{12,1}e^{-a_{1}\omega_{12,1}}\right]+c_{3}\left[\omega_{13,1}e^{-a_{1}\omega_{13,1}}\right]
\end{equation}

and

\[
\left(\frac{\left(M-1\right)}{2M}-\frac{1}{6}\right)\zeta_{k}e^{-z_{1}P_{p}\zeta_{k}\left|g_{1}\right|^{2}a_{k}}+\frac{1}{3}\zeta_{k}e^{-a_{k}z_{2}P_{p}\zeta_{k}\left|g_{1}\right|^{2}}+\frac{\left(\frac{\left(M-1\right)}{2M}-\frac{1}{4}\right)}{\sin^{2}\frac{\pi\left(M-1\right)}{M}}\zeta_{k}e^{-a_{k}z_{3}P_{p}\zeta_{k}\left|g_{1}\right|^{2}}
\]

\begin{equation}
=\left(\frac{\left(M-1\right)}{2M}-\frac{1}{6}\right)\zeta_{1}e^{-z_{1}P_{p}\zeta_{1}\left|g_{1}\right|^{2}a_{1}}+\frac{1}{3}\zeta_{1}e^{-a_{1}z_{2}P_{p}\zeta_{1}\left|g_{1}\right|^{2}}+\frac{\left(\frac{\left(M-1\right)}{2M}-\frac{1}{4}\right)}{\sin^{2}\frac{\pi\left(M-1\right)}{M}}\zeta_{1}e^{-a_{1}z_{3}P_{p}\zeta_{1}\left|g_{1}\right|^{2}}
\end{equation}

\noindent From this expression we can notice that, for a given $\zeta_{1}$
and $\zeta_{k}$, the equality can be satisfied by 

\begin{equation}
\left(\frac{\left(M-1\right)}{2M}-\frac{1}{6}\right)\zeta_{k}e^{-z_{1}P_{p}\zeta_{k}\left|g_{1}\right|^{2}a_{k}}=\left(\frac{\left(M-1\right)}{2M}-\frac{1}{6}\right)\zeta_{1}e^{-z_{1}P_{p}\zeta_{1}\left|g_{1}\right|^{2}a_{1}}
\end{equation}

\begin{equation}
\frac{1}{3}\zeta_{k}e^{-a_{k}z_{2}P_{p}\zeta_{k}\left|g_{1}\right|^{2}}=\frac{1}{3}\zeta_{1}e^{-a_{1}z_{2}P_{p}\zeta_{1}\left|g_{1}\right|^{2}}
\end{equation}

\begin{equation}
\frac{\left(\frac{\left(M-1\right)}{2M}-\frac{1}{4}\right)}{\sin^{2}\frac{\pi\left(M-1\right)}{M}}\zeta_{k}e^{-a_{k}z_{3}P_{p}\zeta_{k}\left|g_{1}\right|^{2}}=\frac{\left(\frac{\left(M-1\right)}{2M}-\frac{1}{4}\right)}{\sin^{2}\frac{\pi\left(M-1\right)}{M}}\zeta_{1}e^{-a_{1}z_{3}P_{p}\zeta_{1}\left|g_{1}\right|^{2}}
\end{equation}

which can be simplified as 

\begin{equation}
e^{-z_{1}P_{p}\zeta_{k}\left|g_{1}\right|^{2}a_{k}}=\frac{\zeta_{1}}{\zeta_{k}}e^{-z_{1}P_{p}\zeta_{1}\left|g_{1}\right|^{2}a_{1}}\label{eq:41}
\end{equation}

\begin{equation}
e^{-a_{k}z_{2}P_{p}\zeta_{k}\left|g_{1}\right|^{2}}=\frac{\zeta_{1}}{\zeta_{k}}e^{-a_{1}z_{2}P_{p}\zeta_{1}\left|g_{1}\right|^{2}}\label{eq:42}
\end{equation}

\begin{equation}
e^{-a_{k}z_{3}P_{p}\zeta_{k}\left|g_{1}\right|^{2}}=\frac{\zeta_{1}}{\zeta_{k}}e^{-a_{1}z_{3}P_{p}\zeta_{1}\left|g_{1}\right|^{2}}\label{eq:43}
\end{equation}

By taking, $\ln$, to the two sides in (\ref{eq:41}), (\ref{eq:42})
and (\ref{eq:43}), we can get

\begin{equation}
a_{k}=\frac{\zeta_{1}a_{1}}{\zeta_{k}}-\frac{\ln\frac{\zeta_{1}}{\zeta_{k}}}{z_{1}P_{p}\zeta_{k}\left|g_{1}\right|^{2}}\label{eq:47}
\end{equation}

\begin{equation}
a_{k}=\frac{\zeta_{1}a_{1}}{\zeta_{k}}-\frac{\ln\frac{\zeta_{1}}{\zeta_{k}}}{z_{2}P_{p}\zeta_{k}\left|g_{1}\right|^{2}}\label{eq:48-1}
\end{equation}

\begin{equation}
a_{k}=\frac{\zeta_{1}a_{1}}{\zeta_{k}}-\frac{\ln\frac{\zeta_{1}}{\zeta_{k}}}{z_{3}P_{p}\zeta_{k}\left|g_{1}\right|^{2}}\label{eq:49}
\end{equation}

 In the cases when the users have same path-loss, we can obtain $a_{k}=a_{1}$
from the all three equations (\ref{eq:47}), (\ref{eq:48-1}) and
(\ref{eq:49}). At high SNR values the last three expressions (\ref{eq:47}),
(\ref{eq:48-1}) and (\ref{eq:49}) can be reduced to

\begin{equation}
a_{k}=\frac{\zeta_{1}a_{1}}{\zeta_{k}}\label{eq:48}
\end{equation}

Substituting (\ref{eq:48}) into (\ref{eq:34}), we can find

\begin{equation}
\stackrel[k=1]{K}{\sum}a_{k}-1=\stackrel[k=1]{K}{\sum}\frac{\zeta_{1}a_{1}}{\zeta_{k}}-1=0
\end{equation}

\begin{equation}
a_{1}=\frac{1}{\zeta_{1}\stackrel[k=1]{K}{\sum}\frac{1}{\zeta_{k}}}\label{eq:51}
\end{equation}

Finally, substituting (\ref{eq:51}) into (\ref{eq:48}) we can get

\begin{equation}
a_{k}=\frac{\zeta_{1}a_{1}}{\zeta_{k}}=\frac{1}{\zeta_{k}\stackrel[k=1]{K}{\sum}\frac{1}{\zeta_{k}}}\label{eq:53}
\end{equation}

In case the users have same path-loss, $\zeta_{1}=..=\zeta_{k}=..\zeta_{K}$,
(\ref{eq:53}) becomes $a_{k}=\frac{1}{K}$. This means that, under
uniform path loss across the users the Min-Sum power allocation reduces
to EPA.

\subsection{Min-Max SEP}

Min-Max power allocation scheme is a widely adopted as fairness criterion;
thus, the obtained design by Min-Max scheme can provide high performance/fairness
of the weak users. In the following, we study power allocation strategy
to minimize the maximum SEP of the considered system subject to the
sum-power constraint. Accordingly, the Min-Max problem can be formulated
as

\[
\underset{\boldsymbol{a}}{\min}\quad\max\left\{ P_{e,1},...,P_{e,k},...P_{e,K}\right\} 
\]

\begin{equation}
S.t:\stackrel[k=1]{K}{\sum}a_{k}=1,\; a_{k}\geq0
\end{equation}

Since the average SEP, $P_{e,k}$, depends totally on the received
SNR at user $k$, the user who has maximum SEP, $P_{e,\textrm{max}}$,
can be defined as the user who has minimum received SNR, $\gamma_{\textrm{min}}=\min\left\{ \gamma_{1},...,\gamma_{k},....,\gamma_{K}\right\} $.
Therefore, maximum SEP can be calculated by

\begin{equation}
P_{e,\textrm{max}}=\frac{1}{\pi}\stackrel[0]{\frac{\pi\left(M-1\right)}{M}}{\int}\mathcal{M}_{\gamma_{min}}\left(z\right)d\theta\label{eq:55}
\end{equation}

\noindent where $\mathcal{M}_{\gamma_{min}}\left(z\right)$ is the
MGF of the minimum received SNR. In order to find $\mathcal{M}_{\gamma_{min}}\left(z\right)$,
we need to find the CDF and/or PDF of $\gamma_{\textrm{min}}$, which
is the distribution of the minimum of dependent random variables.
The CDF of $\gamma_{\textrm{min}}$ can be derived by \cite{bookorderstatis} 

\begin{equation}
F_{\gamma_{min}}\left(\bar{\gamma}\right)=1-\textrm{Pr}\left(\gamma_{1}>\bar{\gamma},...,\gamma_{k}>\bar{\gamma},....,\gamma_{K}>\bar{\gamma}\right)
\end{equation}

It was shown in \cite{bookorderstatis} that

\begin{equation}
\textrm{Pr}\left(\gamma_{1}>\bar{\gamma},...,\gamma_{k}>\bar{\gamma},....,\gamma_{K}>\bar{\gamma}\right)\geq\stackrel[k=1]{K}{\Pi}\textrm{Pr}\left(\gamma_{k}>\bar{\gamma}\right)=\stackrel[k=1]{K}{\Pi}\left[1-\textrm{Pr}\left(\gamma_{k}<\bar{\gamma}\right)\right]
\end{equation}

Based on this fact we can write the CDF of $\gamma_{\textrm{min}}$
as

\begin{equation}
F_{\gamma_{min}}\left(\bar{\gamma}\right)\leq1-\stackrel[k=1]{K}{\Pi}\left[1-F_{\gamma_{k}}\left(\bar{\gamma}\right)\right]\label{eq:61}
\end{equation}

Let $A_{i}$ be the event that $\gamma_{i}$ is selected, then the
PDF of $\gamma_{\textrm{min}}$ can be written as

\begin{equation}
f_{\gamma_{min}}\left(\bar{\gamma},\, A_{i}\right)\leq f_{\gamma_{i}}\left(\bar{\gamma}\right)\stackrel[k=1,k\neq i]{K}{\Pi}\left[1-F_{\gamma_{k}}\left(\bar{\gamma}\right)\right]
\end{equation}

Now, we can calculate the the MGF of the minimum SNR, $\mathcal{M}_{\gamma_{min}}\left(z\right)$.
The MGF of the minimum received SNR is given by

\begin{equation}
\mathcal{M}_{\gamma_{min}}\left(z\right)=\stackrel[0]{\infty}{\int}e^{-z\bar{\gamma}}f_{\gamma_{min}}\left(\bar{\gamma}\right)d\bar{\gamma}
\end{equation}

Using integration by parts we can find that 

\begin{equation}
\mathcal{M}_{\gamma_{min}}\left(z\right)=1-z\stackrel[0]{\infty}{\int}e^{-z\bar{\gamma}}\left(1-F_{\gamma_{min}}\left(\bar{\gamma}\right)\right)d\bar{\gamma}\label{eq:64}
\end{equation}

Substituting (\ref{eq:61}) into (\ref{eq:64}) we can get

\begin{equation}
\mathcal{M}_{\gamma_{min}}\left(z\right)=1-z\stackrel[0]{\infty}{\int}e^{-z\bar{\gamma}}\left(1-\left(1-\stackrel[k=1]{K}{\Pi}\left[1-F_{\gamma_{k}}\left(\bar{\gamma}\right)\right]\right)\right)d\bar{\gamma}
\end{equation}

The CDF of the received SNR can be re-presented as $F_{\gamma_{k}}\left(\bar{\gamma}\right)=\frac{\varphi\left(d/p,\left(\bar{\gamma}/\alpha_{k}\varrho_{k}\right)^{p}\right)}{\Gamma\left(d/p\right)}$,
where $\varphi\left(.\right)$ is the lower incomplete Gamma function.
Thus,

\begin{equation}
\mathcal{M}_{\gamma_{min}}\left(z\right)=1-z\stackrel[0]{\infty}{\int}e^{-z\bar{\gamma}}\left(\stackrel[k=1]{K}{\Pi}\left[1-\frac{\varphi\left(d/p,\left(\bar{\gamma}/\alpha_{k}\varrho_{k}\right)^{p}\right)}{\Gamma\left(d/p\right)}\right]\right)d\bar{\gamma}
\end{equation}

Applying Gaussian Quadrature rule, the MGF can be written as

\begin{equation}
\mathcal{M}_{\gamma_{min}}\left(z\right)=1-\stackrel[i=1]{n}{\sum}\textrm{H}_{i}\left(\stackrel[k=1]{K}{\Pi}\left[1-\frac{\varphi\left(d/p,\left(\bar{\gamma}_{i}/z\alpha_{k}\,\varrho_{k}\right)^{p}\right)}{\Gamma\left(d/p\right)}\right]\right)\label{eq:65}
\end{equation}

\noindent where $\bar{\gamma}_{i}$ is the $i^{th}$ zero of the Laguerre
polynomials \cite{book2}. Substituting (\ref{eq:65}) into (\ref{eq:55}),
the maximum SEP can be calculated by

\begin{equation}
P_{e,\textrm{max}}=\frac{1}{\pi}\stackrel[0]{\frac{\pi\left(M-1\right)}{M}}{\int}\left(1-\stackrel[i=1]{n}{\sum}\textrm{H}_{i}\left(\stackrel[k=1]{K}{\Pi}\left[1-\frac{\varphi\left(d/p,\left(\bar{\gamma}_{i}/z\alpha_{k}\,\varrho_{k}\right)^{p}\right)}{\Gamma\left(d/p\right)}\right]\right)\right)d\theta
\end{equation}

\noindent where $z=-\frac{\sin^{2}\left(\frac{\pi}{M}\right)}{\sin^{2}\theta}$.
Using the approximation formula in (\ref{eq:29}), the max SEP can
be written as

\[
P_{e,max}=\left(\frac{\Theta}{2\pi}-\frac{1}{6}\right)\left(1-\stackrel[i=1]{n}{\sum}\textrm{H}_{i}\left(\stackrel[k=1]{K}{\Pi}\left[1-\frac{\varphi\left(d/p,\left(\bar{\gamma}_{i}/\sin^{2}\left(\frac{\pi}{M}\right)\alpha_{k}\,\varrho_{k}\right)^{p}\right)}{\Gamma\left(d/p\right)}\right]\right)\right)
\]

\[
+\frac{1}{4}\left(1-\stackrel[i=1]{n}{\sum}\textrm{H}_{i}\left(\stackrel[k=1]{K}{\Pi}\left[1-\frac{\varphi\left(d/p,\left(3\bar{\gamma}_{i}/4\sin^{2}\left(\frac{\pi}{M}\right)\alpha_{k}\,\varrho_{k}\right)^{p}\right)}{\Gamma\left(d/p\right)}\right]\right)\right)
\]

\begin{equation}
+\left(\frac{\Theta}{2\pi}-\frac{1}{4}\right)\left(1-\stackrel[i=1]{n}{\sum}\textrm{H}_{i}\left(\stackrel[k=1]{K}{\Pi}\left[1-\frac{\varphi\left(d/p,\left(\bar{\gamma}_{i}\sin^{2}\Theta/\sin^{2}\left(\frac{\pi}{M}\right)\alpha_{k}\,\varrho_{k}\right)^{p}\right)}{\Gamma\left(d/p\right)}\right]\right)\right)
\end{equation}

which can be simplified to 

\[
P_{e,max}=\left(\frac{\Theta}{\pi}-\frac{1}{6}\right)-\left(\frac{\Theta}{2\pi}-\frac{1}{6}\right)\left(\stackrel[i=1]{n}{\sum}\textrm{H}_{i}\left(\stackrel[k=1]{K}{\Pi}\left[1-\frac{\varphi\left(d/p,\left(\bar{\gamma}_{i}/z\alpha_{k}\,\varrho_{k}\right)^{p}\right)}{\Gamma\left(d/p\right)}\right]\right)\right)
\]

\[
-\frac{1}{4}\left(\stackrel[i=1]{n}{\sum}\textrm{H}_{i}\left(\stackrel[k=1]{K}{\Pi}\left[1-\frac{\varphi\left(d/p,\left(\bar{\gamma}_{i}/z\alpha_{k}\,\varrho_{k}\right)^{p}\right)}{\Gamma\left(d/p\right)}\right]\right)\right)
\]

\begin{equation}
-\left(\frac{\Theta}{2\pi}-\frac{1}{4}\right)\left(\stackrel[i=1]{n}{\sum}\textrm{H}_{i}\left(\stackrel[k=1]{K}{\Pi}\left[1-\frac{\varphi\left(d/p,\left(\bar{\gamma}_{i}/z\alpha_{k}\,\varrho_{k}\right)^{p}\right)}{\Gamma\left(d/p\right)}\right]\right)\right)\label{eq:71}
\end{equation}

\noindent Now, the Min-Max problem can be formulated as

\[
\underset{\boldsymbol{a}}{\min}\quad P_{e,max}
\]

\begin{equation}
S.t:\stackrel[k=1]{K}{\sum}a_{k}=1,\; a_{k}\geq0
\end{equation}

which can be expressed using the approximated SEP formula in (\ref{eq:71})
as

\[
\underset{\boldsymbol{a}}{\min}\quad\left(\frac{\Theta}{2\pi}-\frac{1}{6}\right)\mathcal{M}_{\gamma_{min}}\left(\sin^{2}\left(\frac{\pi}{M}\right)\right)+\frac{1}{4}\mathcal{M}_{\gamma_{min}}\left(\frac{4\sin^{2}\left(\frac{\pi}{M}\right)}{3}\right)+\left(\frac{\Theta}{2\pi}-\frac{1}{4}\right)\mathcal{M}_{\gamma_{min}}\left(\frac{\sin^{2}\left(\frac{\pi}{M}\right)}{\sin^{2}\Theta}\right)
\]

\begin{equation}
S.t:\stackrel[k=1]{K}{\sum}a_{k}=1,\; a_{k}\geq0
\end{equation}

and

\[
\underset{\boldsymbol{a}}{\min}\quad\left(\frac{\Theta}{\pi}-\frac{1}{6}\right)-c_{1}\left(\stackrel[i=1]{n}{\sum}\textrm{H}_{i}\left(\stackrel[k=1]{K}{\Pi}\left[1-\frac{\varphi\left(d/p,\left(\bar{\gamma}_{i}/z_{1}a_{k}P_{p}\zeta_{k}\,\varrho_{k}\right)^{p}\right)}{\Gamma\left(d/p\right)}\right]\right)\right)
\]

\[
-c_{2}\left(\stackrel[i=1]{n}{\sum}\textrm{H}_{i}\left(\stackrel[k=1]{K}{\Pi}\left[1-\frac{\varphi\left(d/p,\left(\bar{\gamma}_{i}/z_{2}a_{k}P_{p}\zeta_{k}\,\varrho_{k}\right)^{p}\right)}{\Gamma\left(d/p\right)}\right]\right)\right)
\]

\[
-c_{3}\left(\stackrel[i=1]{n}{\sum}\textrm{H}_{i}\left(\stackrel[k=1]{K}{\Pi}\left[1-\frac{\varphi\left(d/p,\left(\bar{\gamma}_{i}/z_{3}a_{k}P_{p}\zeta_{k}\,\varrho_{k}\right)^{p}\right)}{\Gamma\left(d/p\right)}\right]\right)\right)
\]

\begin{equation}
S.t:\stackrel[k=1]{K}{\sum}a_{k}=1,\; a_{k}\geq0
\end{equation}

\noindent $c_{1}=\left(\frac{\Theta}{2\pi}-\frac{1}{6}\right)$,$c_{2}=\frac{1}{4}$,
$c_{3}=\left(\frac{\Theta}{2\pi}-\frac{1}{4}\right)$, $z_{1}=\sin^{2}\left(\frac{\pi}{M}\right)$,
$z_{2}=\frac{4\sin^{2}\left(\frac{\pi}{M}\right)}{3}$ and $z_{3}=\frac{\sin^{2}\left(\frac{\pi}{M}\right)}{\sin^{2}\frac{\pi\left(M-1\right)}{M}}$.
Considering the first-order Laguerre polynomial, we can get

\[
\underset{\boldsymbol{a}}{\min}\quad\left(\frac{\Theta}{\pi}-\frac{1}{6}\right)-c_{1}\left(\textrm{H}_{1}\left(\stackrel[k=1]{K}{\Pi}\left[1-\frac{\varphi\left(d/p,\left(\bar{\gamma}_{1}/z_{1}a_{k}P_{p}\zeta_{k}\,\varrho_{k}\right)^{p}\right)}{\Gamma\left(d/p\right)}\right]\right)\right)
\]

\[
-c_{2}\left(\textrm{H}_{1}\left(\stackrel[k=1]{K}{\Pi}\left[1-\frac{\varphi\left(d/p,\left(\bar{\gamma}_{1}/z_{2}a_{k}P_{p}\zeta_{k}\,\varrho_{k}\right)^{p}\right)}{\Gamma\left(d/p\right)}\right]\right)\right)-c_{3}\left(\textrm{H}_{1}\left(\stackrel[k=1]{K}{\Pi}\left[1-\frac{\varphi\left(d/p,\left(\bar{\gamma}_{1}/z_{3}a_{k}P_{p}\zeta_{k}\,\varrho_{k}\right)^{p}\right)}{\Gamma\left(d/p\right)}\right]\right)\right)
\]

\begin{equation}
S.t:\stackrel[k=1]{K}{\sum}a_{k}=1,\; a_{k}\geq0\label{eq:75}
\end{equation}

The lower incomplete gamma function is given by 

\begin{equation}
\varphi\left(s,x\right)=\stackrel[0]{x}{\int}r^{s-1}e^{-r}\, dr
\end{equation}

It is noted that, the second derivation of the lower incomplete gamma
function can be found as, $\frac{\partial^{2}}{\partial x}\varphi=\left(s-x-1\right)e^{-x}x^{s-2}$.
Since the convexity requires that the second derivative is not negative,
this condition is satisfied of the lower incomplete gamma function
only if $s>x-1$, which means that $\frac{d}{p}>\left(\frac{\bar{\gamma}_{1}}{za_{k}P_{p}\zeta_{k}\,\varrho_{k}}\right)^{p}-1$,
and $2N>\sqrt{\left(\frac{\bar{\gamma}_{1}}{za_{k}P_{p}\zeta_{k}\,\varrho_{k}}\right)}-1$.
As we can see, this optimization problem in (\ref{eq:75}) is hard
to solve numerically, and any closed form solution is hard if not
impossible to find. However, some numerical software tools such as
Mathematica, can be used to solve this problem and thus the optimal
power allocation can be obtained.

\section{Throughput and Power Efficiency\label{sec:Throughput-and-Power}}

In this section we consider the throughput and power efficiency of
the CI precoding in MU-MIMO systems. As the CI has been proposed to
enhance the received SNR, it is important to consider and investigate
the throughput performance of the CI technique. The throughput can
be calculated using the following definition \cite{throuput,CI4}

\begin{equation}
\tau=\left(1-P_{B}\right)\times c\times F\times K\label{eq:80}
\end{equation}

\noindent where $P_{B}$ is the block error rate, $c=\log_{2}\left(M\right)$
is the bit per symbol and $F$ is the block length. The transmission
in communication systems is generally based on sending blocks of $\mathscr{N}=c\times F$
sequential bits, where each block of $\mathscr{N}$ bits might represent
sub or complete a user message. Therefore, the performance of such
systems depends essentially on the probability of errors in each block.
For coherent PSK modulation and in white Gaussian noise environment,
the errors in each block are Binomially distributed. Thus, the probability
of $q$ errors in one block can be expressed as \cite{BLER2,BLER1,BLER3,BLER6,BLER7} 

\begin{equation}
\textrm{Pr}\left(q,\mathscr{N}\right)=\left(\begin{array}{c}
\mathscr{N}\\
q
\end{array}\right)P_{b}^{q}\left(1-P_{b}\right)^{\mathscr{N}-q}\label{eq:81}
\end{equation}

\noindent where $P_{b}$ is the bit error probability (BEP) and can
be calculated using the SEP derivation in Section \ref{sec:Average-Symbol-Error}.
Consequently, the $P_{B}$ in fading channels for a block of $\mathscr{N}$
bits capable of correcting $Q$ errors can be written as \cite{BLER2,BLER1,BLER3,BLER6,BLER7} 

\begin{equation}
P_{B}=1-\stackrel[q=0]{Q}{\sum}\left(\begin{array}{c}
\mathscr{N}\\
q
\end{array}\right)P_{b}^{q}\left(1-P_{b}\right)^{\mathscr{N}-q}\label{eq:82}
\end{equation}

In case the receiver employs only error detection technique, a block
is received correctly only if all $\mathscr{N}$ bits in the block
are received successfully. Therefore, the overall system performance
of such systems relies on the probability of occurrence of one or
more bit errors in a block, i.e., $\textrm{Pr}\left(0,\mathscr{N}\right)$.
On the other hand, if the receiver employs error-correction techniques
which are able to correct up to $Q$ errors in a block, the system
performance is dominated by the probability of occurrence of more
than $Q$ errors in a block, i.e., $\textrm{Pr}\left(Q,\mathscr{N}\right)$.
In case when $Q=0$ and $\mathscr{N}=1$, $P_{B}$ becomes the BEP
\cite{BLER2,BLER1,BLER6,BLER7} . This definition of the $P_{B}$
has been widely studied in literature, for instance \cite{BLER2,BLER1,BLER6,BLER7}
. For simplicity and mathematical tractability we employ the below
approximate expression to derive the BEP from our SEP derivation above\cite[(8.119)]{BER1,salimbook} 

\begin{equation}
P_{b}\backsimeq\frac{2}{\max\left(\log_{2}M,\,2\right)}\stackrel[i=1]{\max\left(\frac{M}{4},\,1\right)}{\sum}\frac{1}{\pi}\times\stackrel[0]{\pi/2}{\int}\mathcal{M}_{\gamma}\left(-\frac{\log_{2}M}{\sin^{2}\theta}\sin^{2}\frac{\left(2i-1\right)\pi}{M}\right)d\theta\label{eq:83-1}
\end{equation}

Substituting (\ref{eq:16}) into (\ref{eq:83-1}) we can get

\begin{equation}
P_{b}\backsimeq\frac{2}{\max\left(\log_{2}M,\,2\right)}\stackrel[i=1]{\max\left(\frac{M}{4},\,1\right)}{\sum}\frac{1}{\pi}\times\stackrel[i=1]{n}{\sum}\frac{\textrm{H}_{i}\left(\frac{p}{a^{d}}\right)\left(\gamma_{i}\right)^{d-1}e^{-\left(\frac{\gamma_{i}}{a}\right)^{p}}}{\Gamma\left(\frac{d}{p}\right)}\stackrel[0]{\pi/2}{\int}e^{-\left(-\frac{\log_{2}M}{\sin^{2}\theta}\sin^{2}\frac{\left(2i-1\right)\pi}{M}-1\right)\gamma_{i}}d\theta
\end{equation}

\noindent which can be found as

\[
P_{b}\backsimeq\frac{2}{\max\left(\log_{2}M,\,2\right)}\stackrel[i=1]{\max\left(\frac{M}{4},\,1\right)}{\sum}\frac{1}{\pi}
\]

\begin{equation}
\times\stackrel[i=1]{n}{\sum}\textrm{H}_{i}\left(\frac{\left(\frac{p}{a^{d}}\right)\left(\gamma_{i}\right)^{d-1}e^{-\left(\frac{\gamma_{i}}{a}\right)^{p}}}{\Gamma\left(\frac{d}{p}\right)}\right)\left(\frac{\pi}{2}e^{\gamma_{i}}\textrm{Erfc}\left(\sqrt{-\log_{2}M\times\sin^{2}\frac{\left(2i-1\right)\pi}{M}\gamma_{i}}\right)\right)\label{eq:86}
\end{equation}

Finally, substituting the BEP expression in (\ref{eq:86}) into (\ref{eq:82})
and then into (\ref{eq:80}) we can find the system throughput. Similarly,
in the communication systems where the decoding depends on the symbol
error, the $P_{B}$ can be evaluated using the SEP. In this case we
can define $\mathscr{N}$ as number of symbols in each block and $Q$
as number of symbol errors, thus $P_{B}$ can be evaluated by replacing
$P_{b}$ with $P_{e}$ in (\ref{eq:82}) \cite{BLERSER,BLER3}. In
the special case when $Q=0$ and $\mathscr{N}=1$, $P_{B}$ becomes
the SEP. Hence, the throughput in this case can be calculated as in
the following expression \cite{BLERSER,BLER3} 

\begin{equation}
\tau=\left(\stackrel[q=0]{Q}{\sum}\left(\begin{array}{c}
\mathscr{N}\\
q
\end{array}\right)P_{e,k}^{q}\left(1-P_{e.k}\right)^{\mathscr{N}-q}\right)\times\log_{2}\left(M\right)\times\mathscr{N}\times K
\end{equation}

\noindent where the exact $P_{e.k}$ is given in (\ref{eq:25}), and
the approximate $P_{e.k}$ is given (\ref{eq:29-1}).

The derived expression of the throughput can be used now to calculate
the power efficiency (PE). The power efficiency combines both the
throughput with the power consumption at the BS, and can be expressed
as \cite{CI4}

\begin{equation}
\textrm{PE}=\frac{\tau}{P_{tot}}
\end{equation}

\noindent where $P_{tot}$ is the total power consumed during the
transmission. In practical systems, the total power can be calculated
by \cite{powerconsmain,powercons60,powercons62}

\begin{equation}
P_{tot}=\frac{P_{PA}+P_{RF}+P_{DS}}{\left(1-\varsigma_{DC}\right)\left(1-\varsigma_{MS}\right)\left(1-\varsigma_{cool}\right)}
\end{equation}

\noindent where $\varsigma_{DC}$, $\varsigma_{MS}$ and $\varsigma_{cool}$
represent the losses of the DC-DC supply, main power supply and the
active cooling, respectively \cite{powerconsmain,powercons60}. In
addition, $P_{PA}$ is the average power consumption of the amplifiers
and given by $P_{PA}=\frac{P_{P}}{\eta_{pa}}$, where $\eta_{pa}$
is the efficiency of the power amplifiers. Furthermore, $P_{RF}$
is the power consumption of the other electronic components in the
RF chains, and can be written as $P_{RF}=N\left(P_{D}+P_{m}+P_{f}\right)+P_{sy}$,
where $P_{D}$, $P_{m}$ and $P_{f}$ are the power consumption of
the digital-to-analog converters, signal mixers and filters, respectively,
while $P_{sy}$ is the power consumption at the frequency synthesizer.
Moreover, $P_{DS}$ is the power consumed by the digital signal processor
\cite{powerconsmain,powercons60,powercons62}.

\section{Numerical Results \label{sec:Numerical-Results}}

This section presents simulation and numerical results of the derived
expressions in this paper. Monte-Carlo simulations are performed with
$10^{6}$ independent trials. It is assumed that, the users have same
noise power, $\sigma^{2}$, and thus the transmit SNR ($\eta_{t}$
) is defined as $\eta_{t}=\frac{P_{p}}{\sigma^{2}}$. In addition,
the path-loss exponent in this section is chosen to be $m=2.7$. 

\begin{figure}
\noindent \begin{centering}
\subfloat[\label{fig:2-a}The CDF of the received SNR for different transmit
SNR values,$\eta_{t}$, when $N=K=4$, and $\mathbf{u}=\mathbf{a}_{k}$.]{\begin{centering}
\includegraphics[scale=0.35]{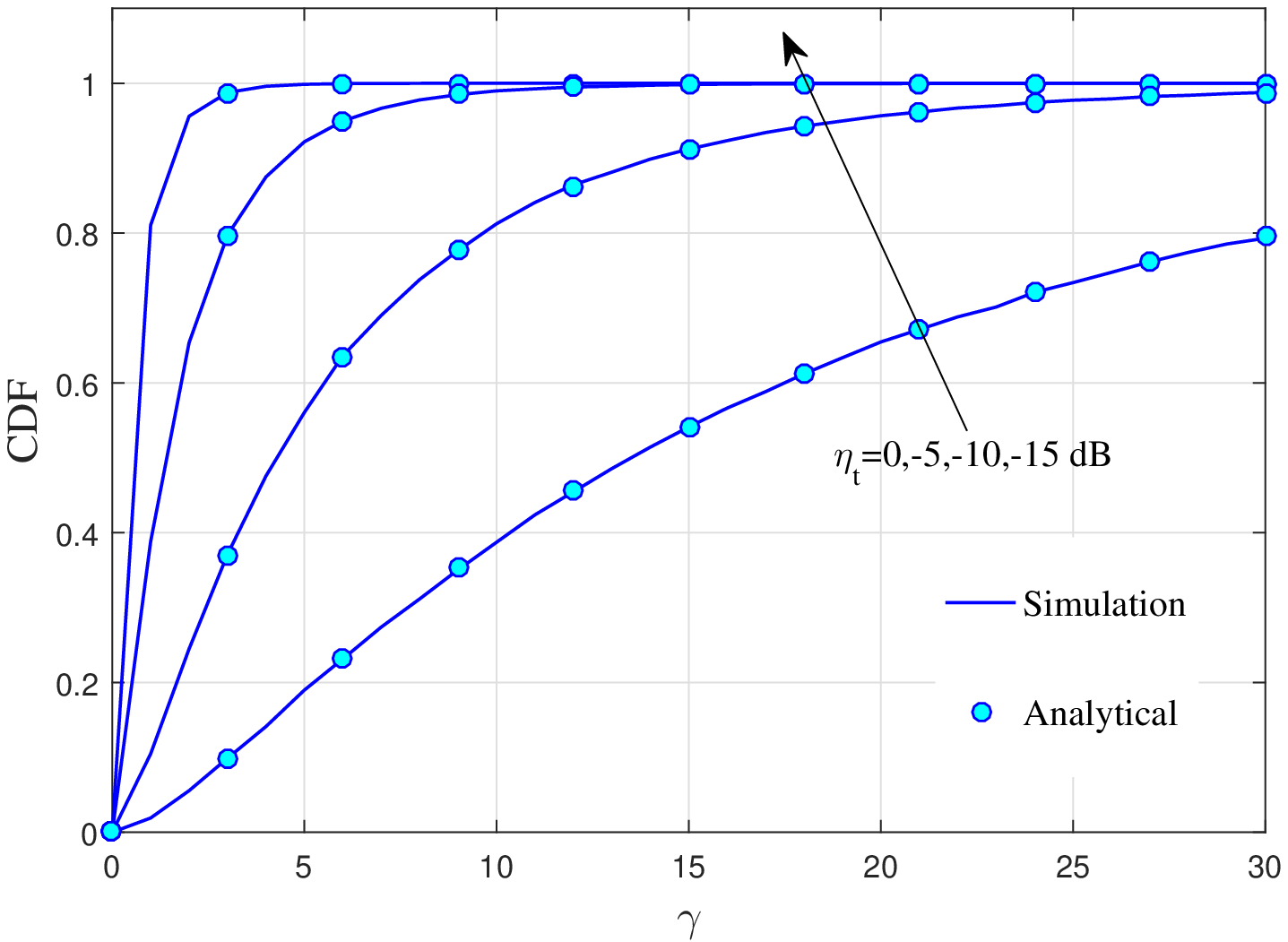}
\par\end{centering}

}\subfloat[\label{fig:2-b}The CDF of the received SNR for different transmit
SNR values,$\eta_{t}$, when $N=6,\, K=4$, and $\mathbf{u}=\mathbf{a}_{k}$.]{\noindent \begin{centering}
\includegraphics[scale=0.35]{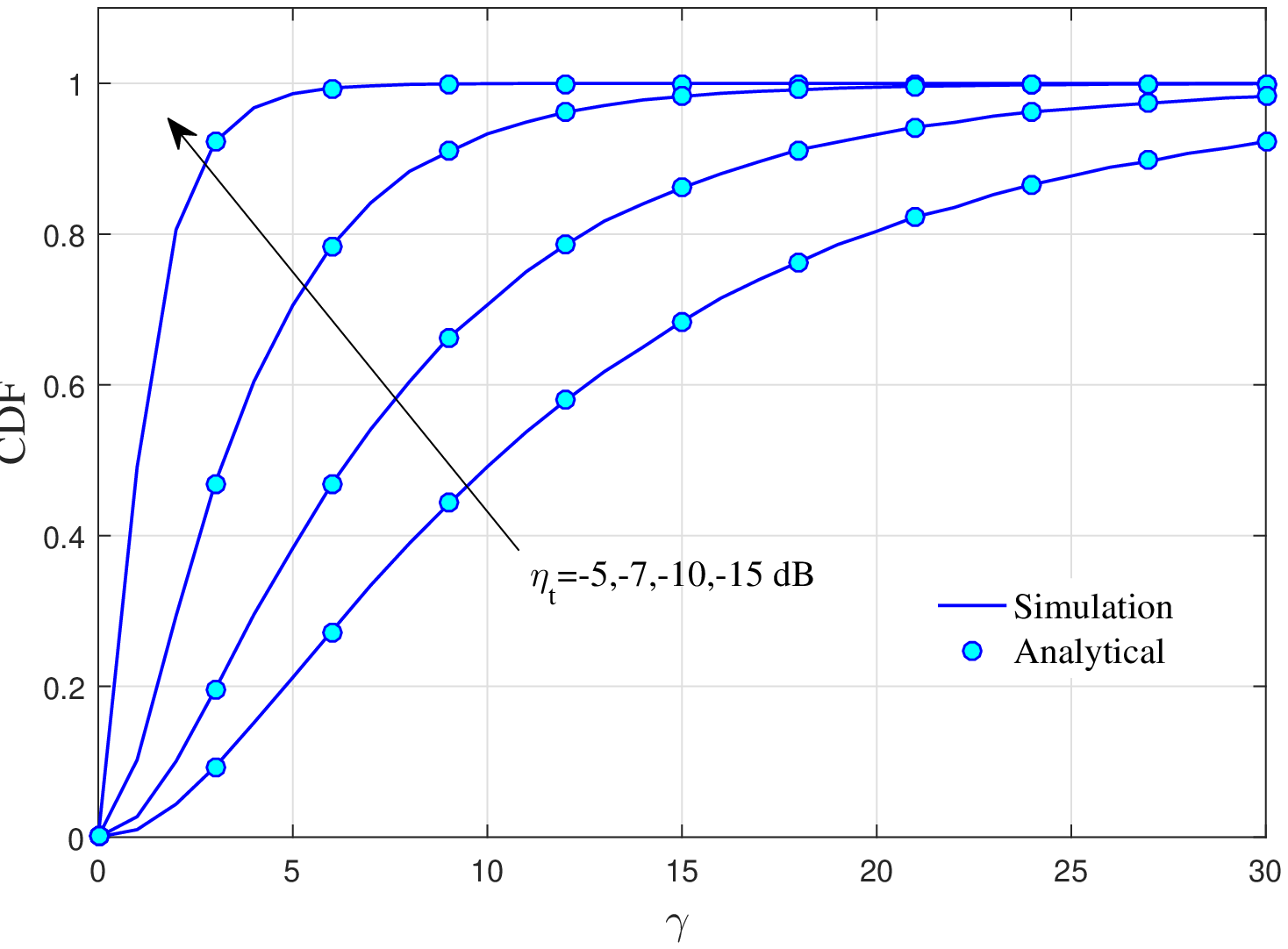}
\par\end{centering}

}\subfloat[\label{fig:2-c}The CDF of the received SNR for different transmit
SNR values,$\eta_{t}$, when $N=K=2$, and $\mathbf{u}=\frac{1}{K}\mathbf{1}$.]{\noindent \begin{centering}
\includegraphics[scale=0.35]{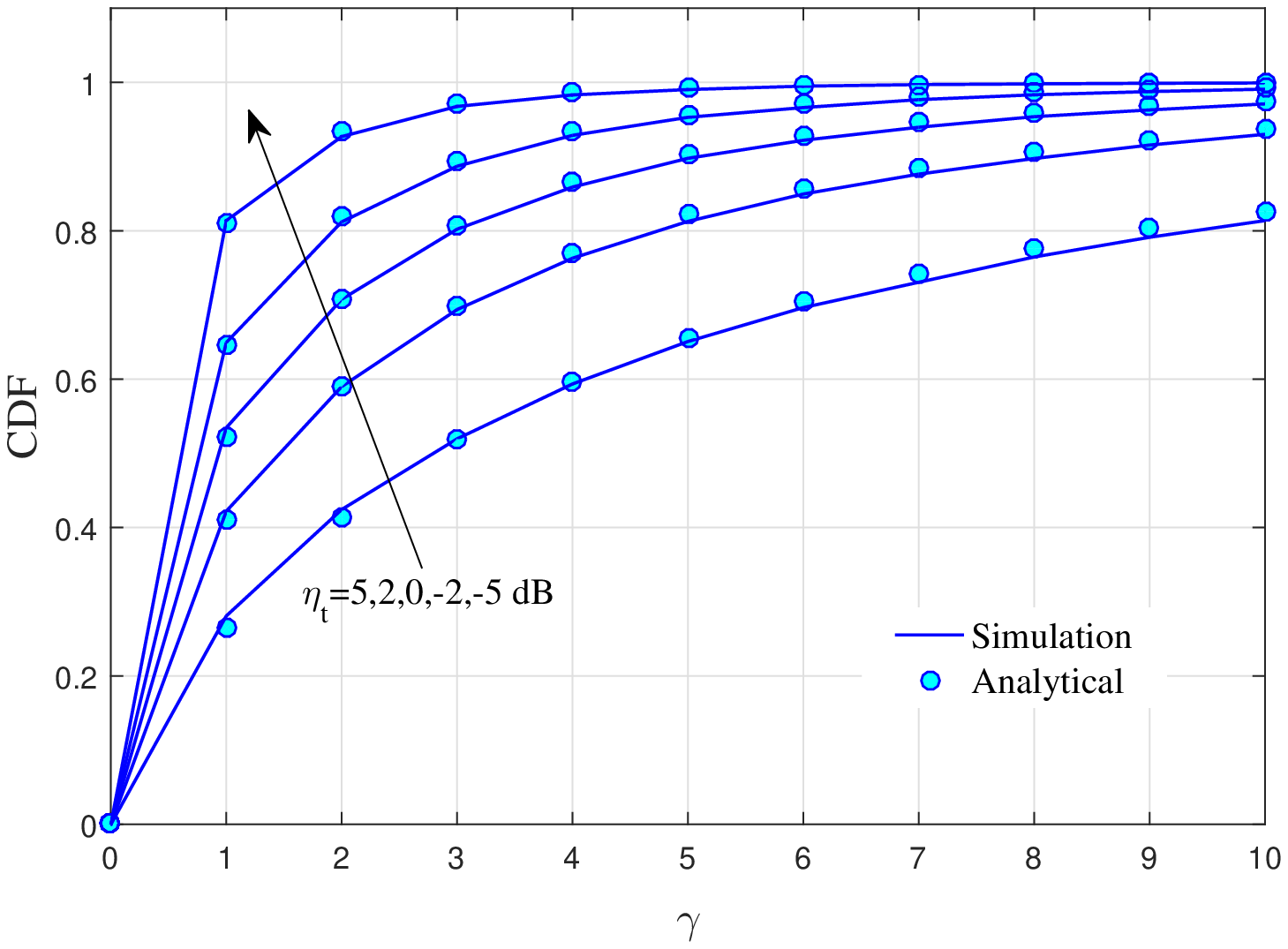}
\par\end{centering}

}
\par\end{centering}

\begin{centering}
\subfloat[\label{fig:2-d}The CDF of the received SNR for different transmit
SNR values,$\eta_{t}$, when $N=K=3$, and $\mathbf{u}=\frac{1}{K}\mathbf{1}$.]{\noindent \begin{centering}
\includegraphics[scale=0.35]{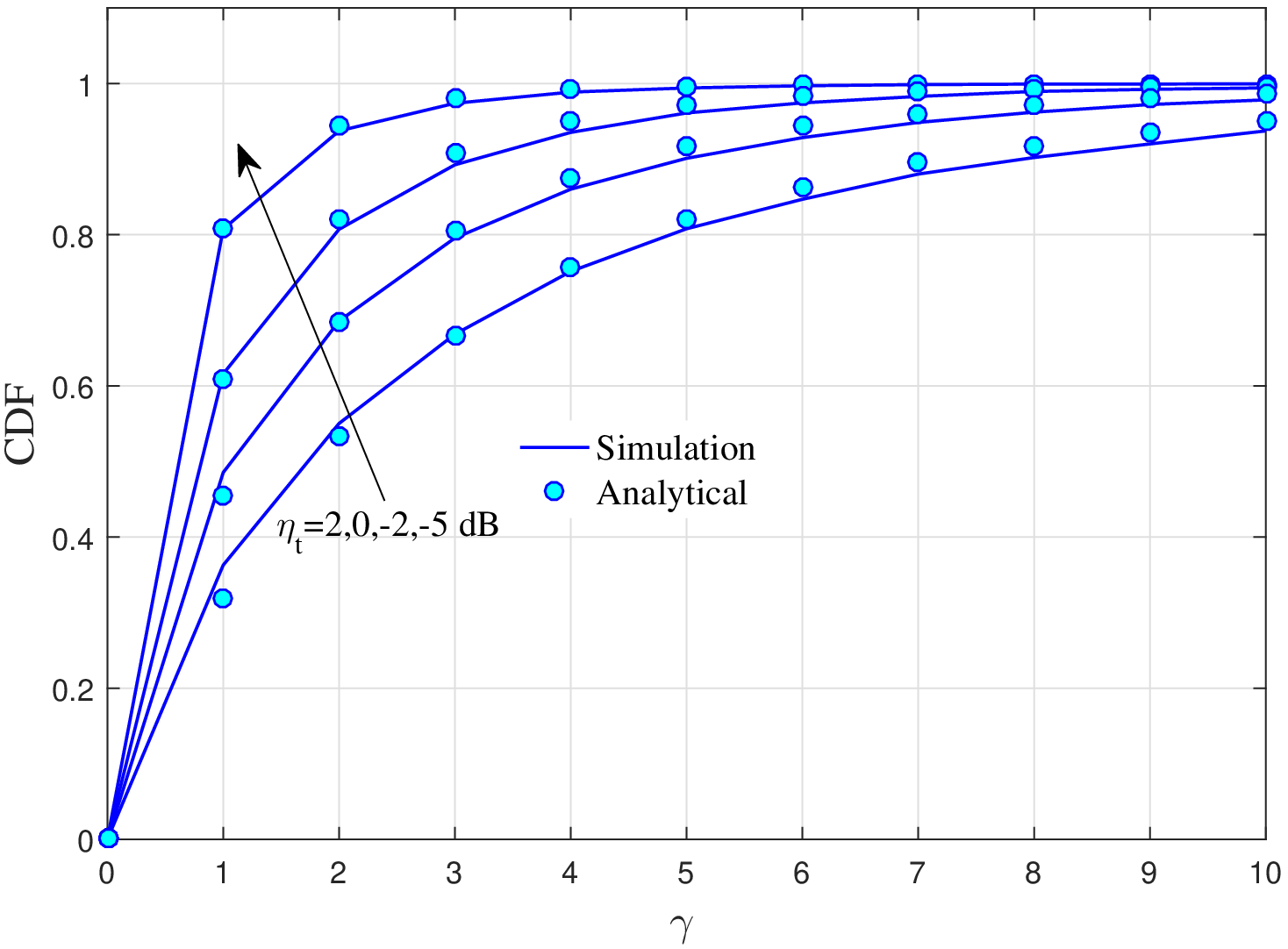}
\par\end{centering}

}\subfloat[\label{fig:2-e}The CDF of the received SNR for different transmit
SNR values,$\eta_{t}$, when $N=8,\, K=5$, and $\mathbf{u}=\frac{1}{K}\mathbf{1}$.]{\noindent \begin{centering}
\includegraphics[scale=0.35]{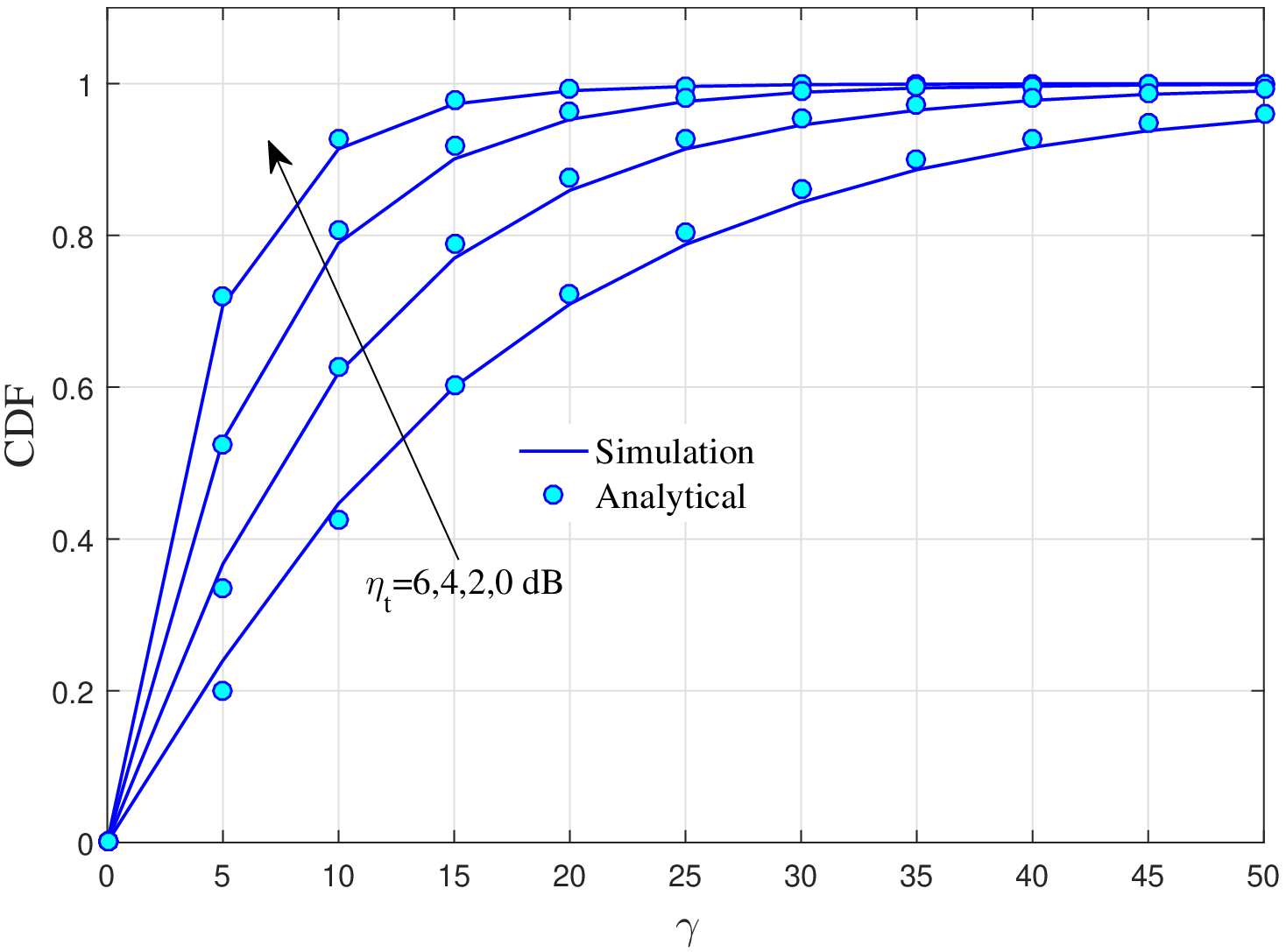}
\par\end{centering}

}\subfloat[\label{fig:2-f}The CDF of the received SNR for different transmit
SNR values,$\eta_{t}$, when $N=K=3$, and $u_{k}=0.2$.]{\noindent \centering{}\includegraphics[scale=0.35]{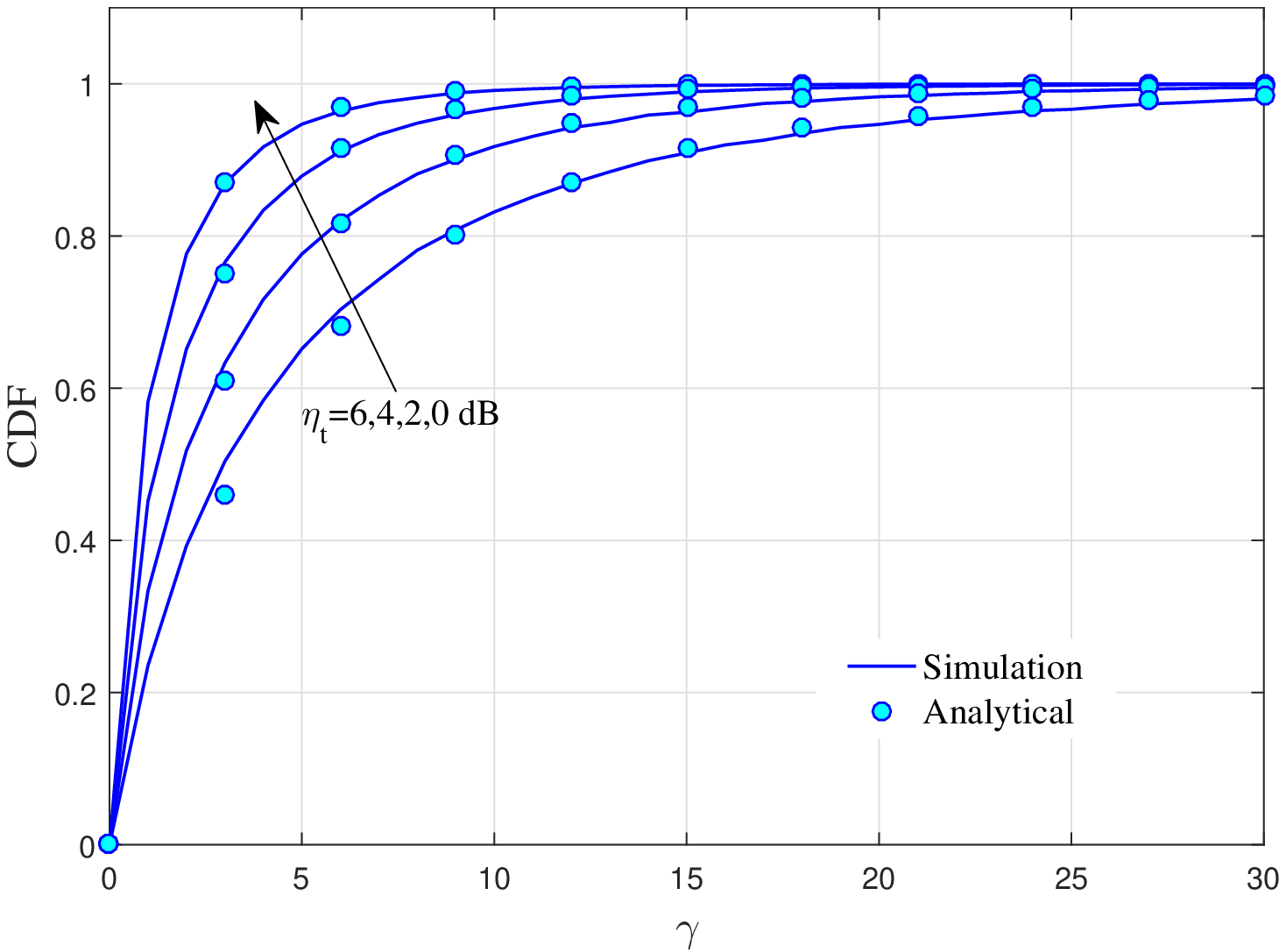}
}
\par\end{centering}

\protect\caption{\label{fig:2}The CDF of the received SNR for different values of
the transmit SNR, $\eta_{t}$, number of users $K$, number of BS
antennas $N$ and $\mathbf{u}$.}
\end{figure}

Firstly, in Fig. \ref{fig:2} we plot the CDF of the received SNR
at the $k^{th}$ user for different values of the transmit SNR, $\eta_{t}$,
number of users, $K$, number of BS antennas, $N$, and the vector
$\mathbf{u}$. The analytical and simulation results are in well agreement,
which confirms the accuracy of the distribution considered in Section
(\ref{sec:MGF-DERIVATION}). In addition, from these results it is
clear that, the values of the elements of $\mathbf{u}$ have impact
on the CDF and thus on the system performance in general. In this
regard it is noted that, user $k$ can achieve the optimal performance
when $u_{k}=1$, which is the case presented in Figs. \ref{fig:2-a}
and \ref{fig:2-b}. Furthermore, the CDF of the received SNR for different
values of $N$ and $K$ when the elements of $\mathbf{u}$ have same
value, $u_{k}=\frac{1}{K}$, are presented in Figs. \ref{fig:2-c},
\ref{fig:2-d}, and \ref{fig:2-e} and when $u_{k}$ has the smallest
value is presented in Fig. \ref{fig:2-f}. In all these cases the
variance of the received SNR will be reduced by the value of $u_{k}$,
and thus smaller value of $u_{k}$ will result in poorer/weaker performance/SNR
of user $k$ in the system. Finally, it is worthy mentioning that,
the results presented in Fig. \ref{fig:2}, can be used also to present
the outage probability of CI precoding technique. The outage probability
is the probability that the received SNR, $\gamma_{k}$, falls below
an acceptable threshold value, $\gamma_{th}$. Therefore, we can obtain
the outage probability of CI precoding by replacing $\gamma$ with
$\gamma_{th}$. 

\begin{figure}
\begin{centering}
\subfloat[\label{fig:3-a}Average received SNR versus transmit SNR,$\eta_{t}$,
for different values of $N$ when $K=4$. ]{\begin{centering}
\includegraphics[scale=0.5]{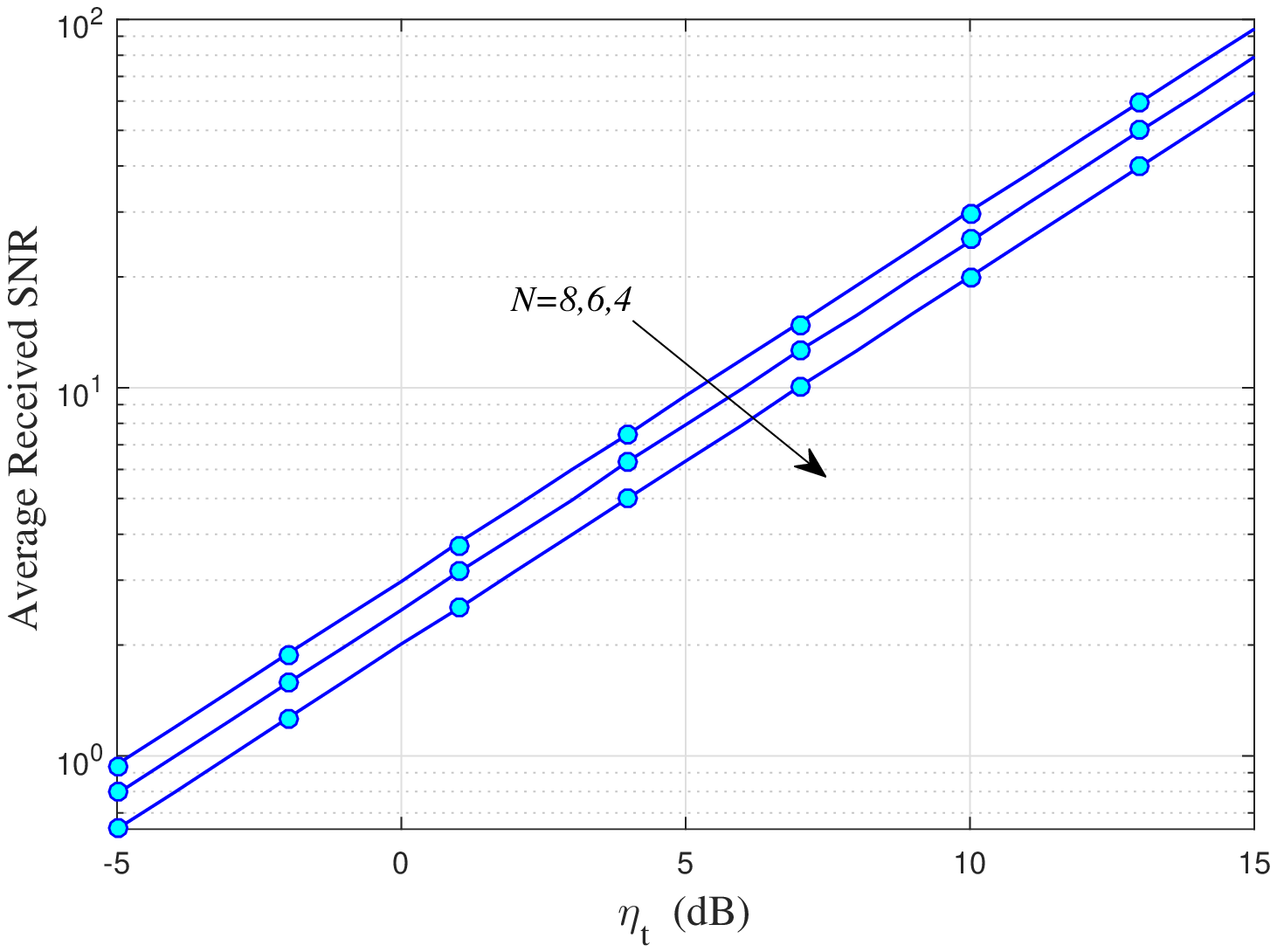}
\par\end{centering}

}\subfloat[\label{fig:3-b}Average received SNR versus transmit SNR,$\eta_{t}$
, for different values of $N$ when $K=2$. ]{\begin{centering}
\includegraphics[scale=0.5]{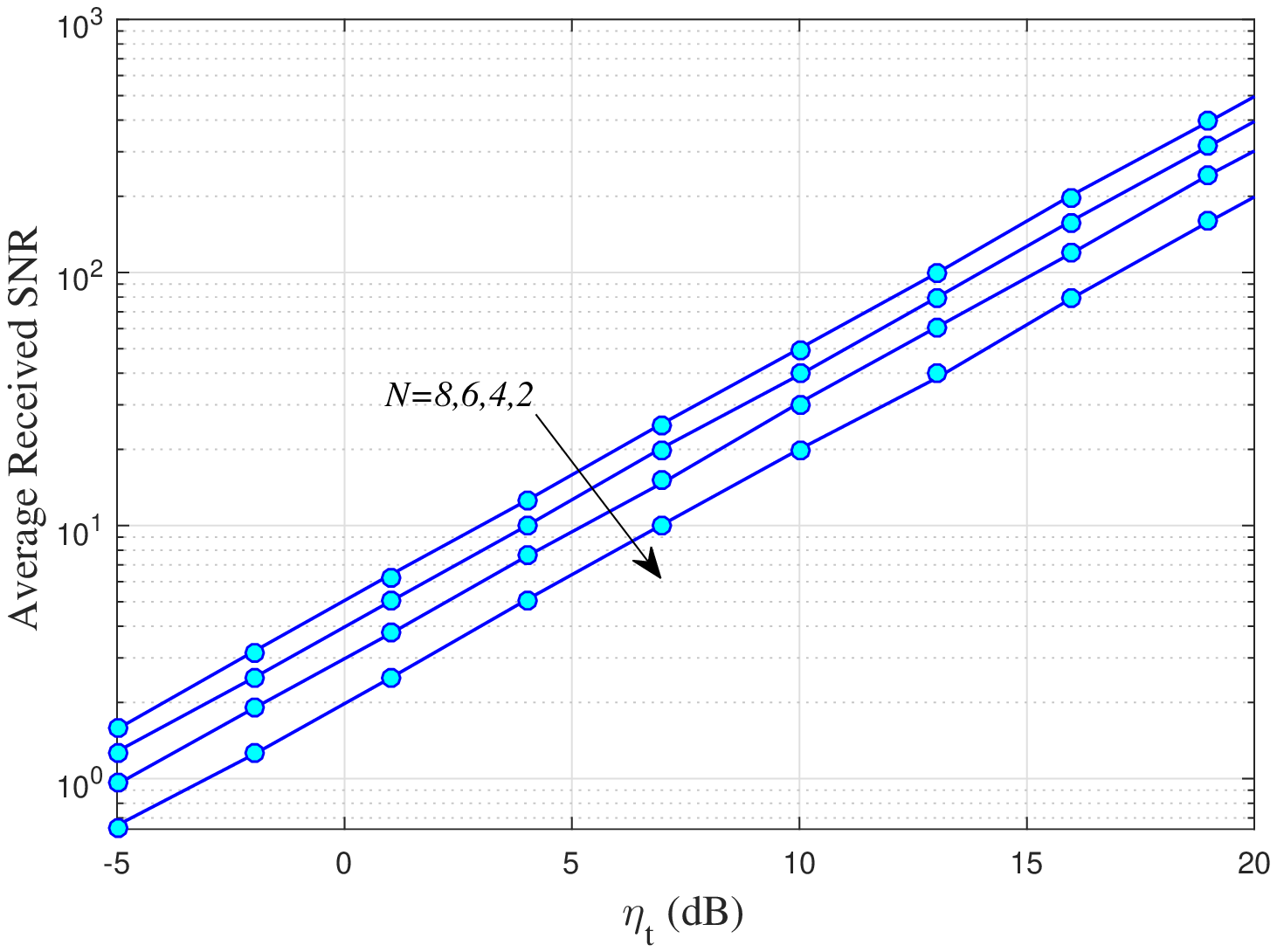}
\par\end{centering}

}
\par\end{centering}

\protect\caption{\label{fig:3}Average received SNR versus transmit SNR, $\eta_{t}$,
for different values of $N$ and $K$.}
\end{figure}

Fig. \ref{fig:3}, illustrates the average received SNR versus the
transmit SNR, $\eta_{t}$, for different values of $N$ and $K$.
Fig. \ref{fig:3-a}, presents the average received SNR when $K=4$
and Fig. \ref{fig:3-b}, shows the average received SNR when $K=2$.
The good matching between the analytical and simulation results confirms
the derived expressions in Section (\ref{sub:Average-SINR}). Generally
and as anticipated, increasing the transmit SNR, number of antennas
and/or number of users lead to enhance the average received SNR. In
addition, the gain attained by increasing number of the antennas is
almost fixed with the transmit SNR in the all considered scenarios.

\begin{figure}
\noindent \begin{centering}
\subfloat[\label{fig:4a}SEP versus transmit SNR, $\eta_{t}$, with different
types of input, when $N=K=4$.]{\noindent \begin{centering}
\includegraphics[scale=0.5]{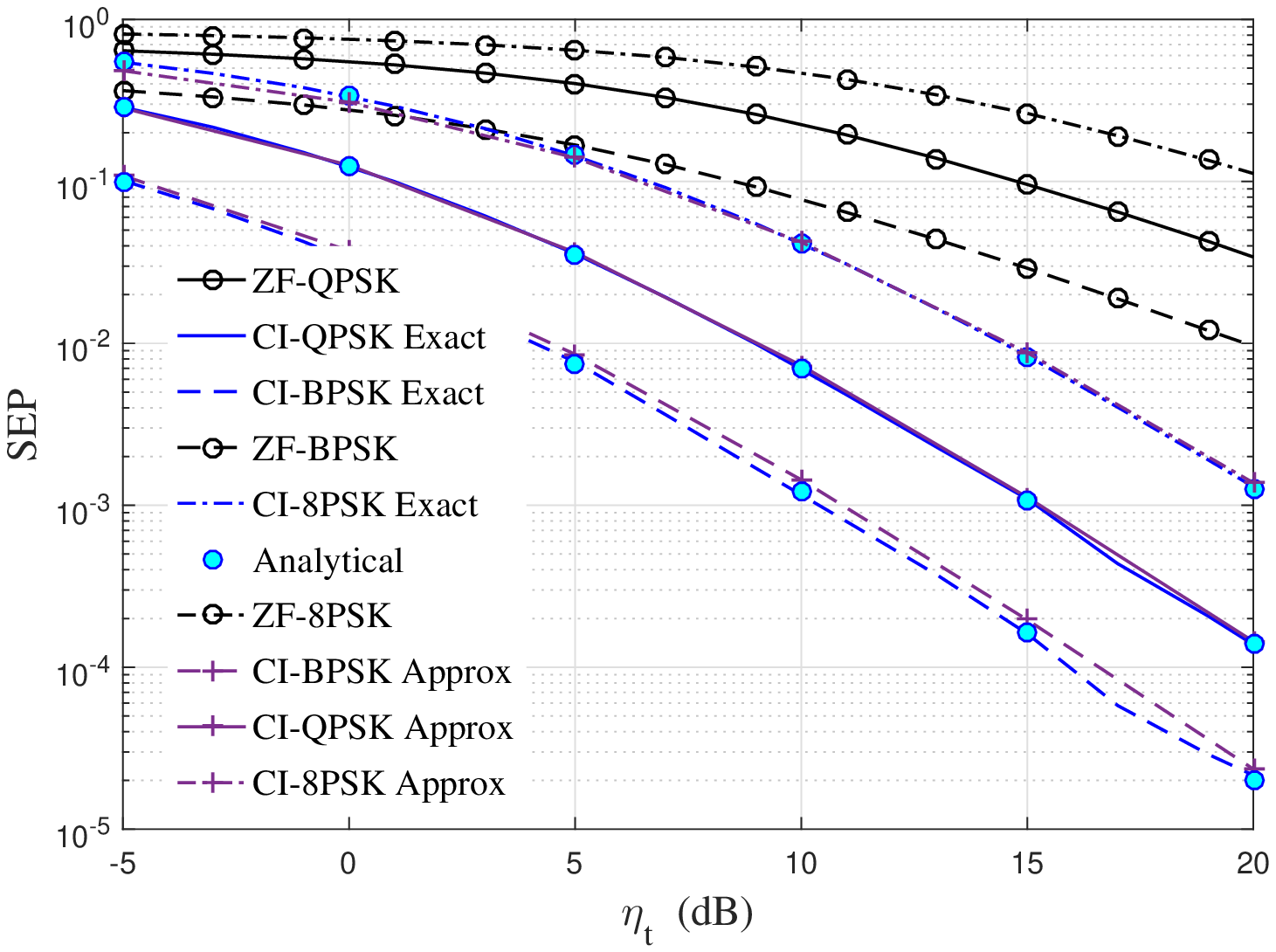}
\par\end{centering}

}\subfloat[\label{fig:4b}SEP versus transmit SNR, $\eta_{t}$, with different
types of input, when $N=6,\, K=4$.]{\begin{centering}
\includegraphics[scale=0.5]{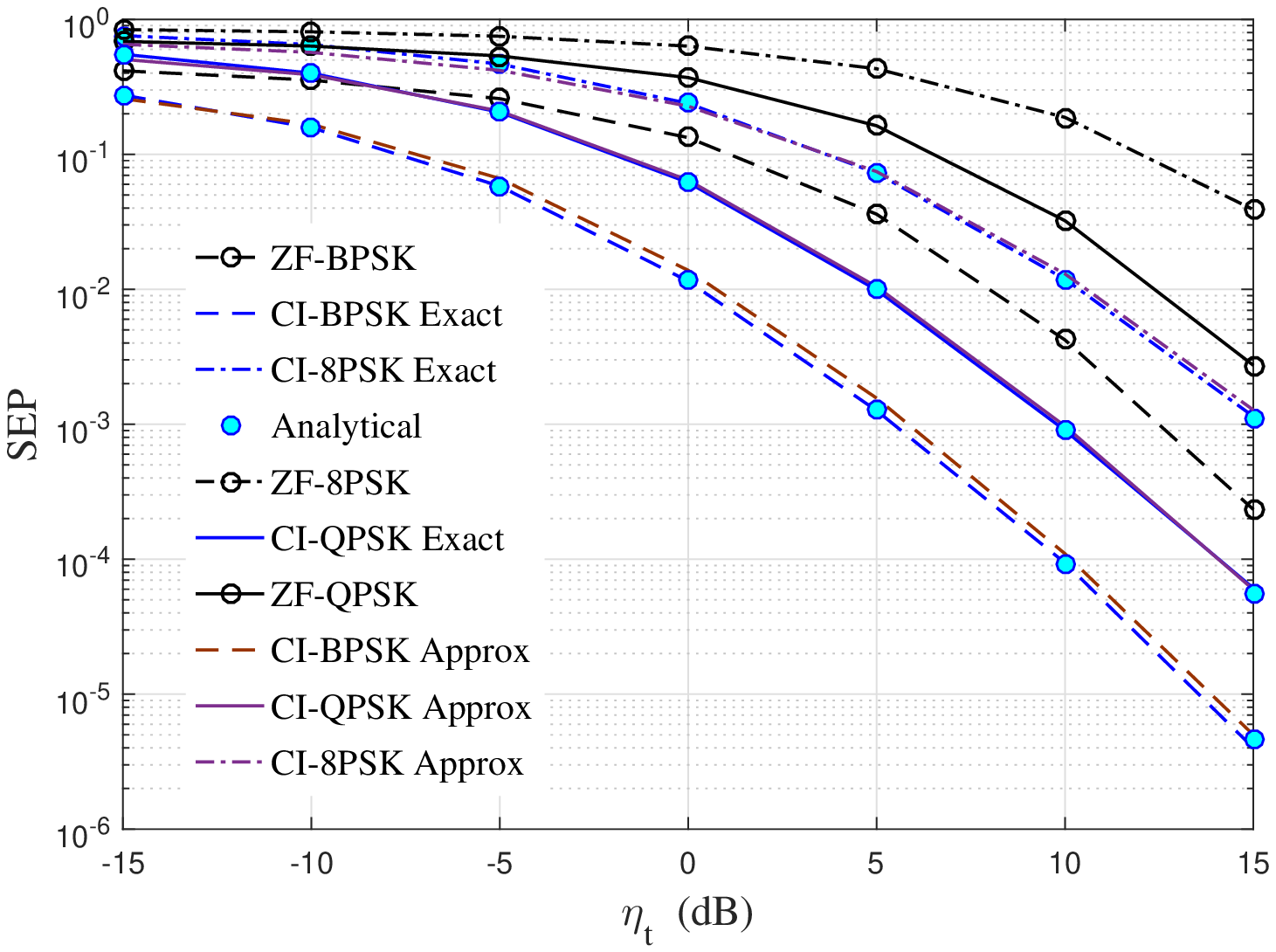}
\par\end{centering}

}
\par\end{centering}

\noindent \centering{}\protect\caption{\label{fig:4}SEP versus transmit SNR for various input types, when
$N=4,6\textrm{ and }K=4$.}
\end{figure}

\begin{figure}
\noindent \begin{centering}
\subfloat[\label{fig:5a}SEP versus transmit SNR, $\eta_{t}$, with different
types of input, when $N=K=6$.]{\begin{centering}
\includegraphics[scale=0.5]{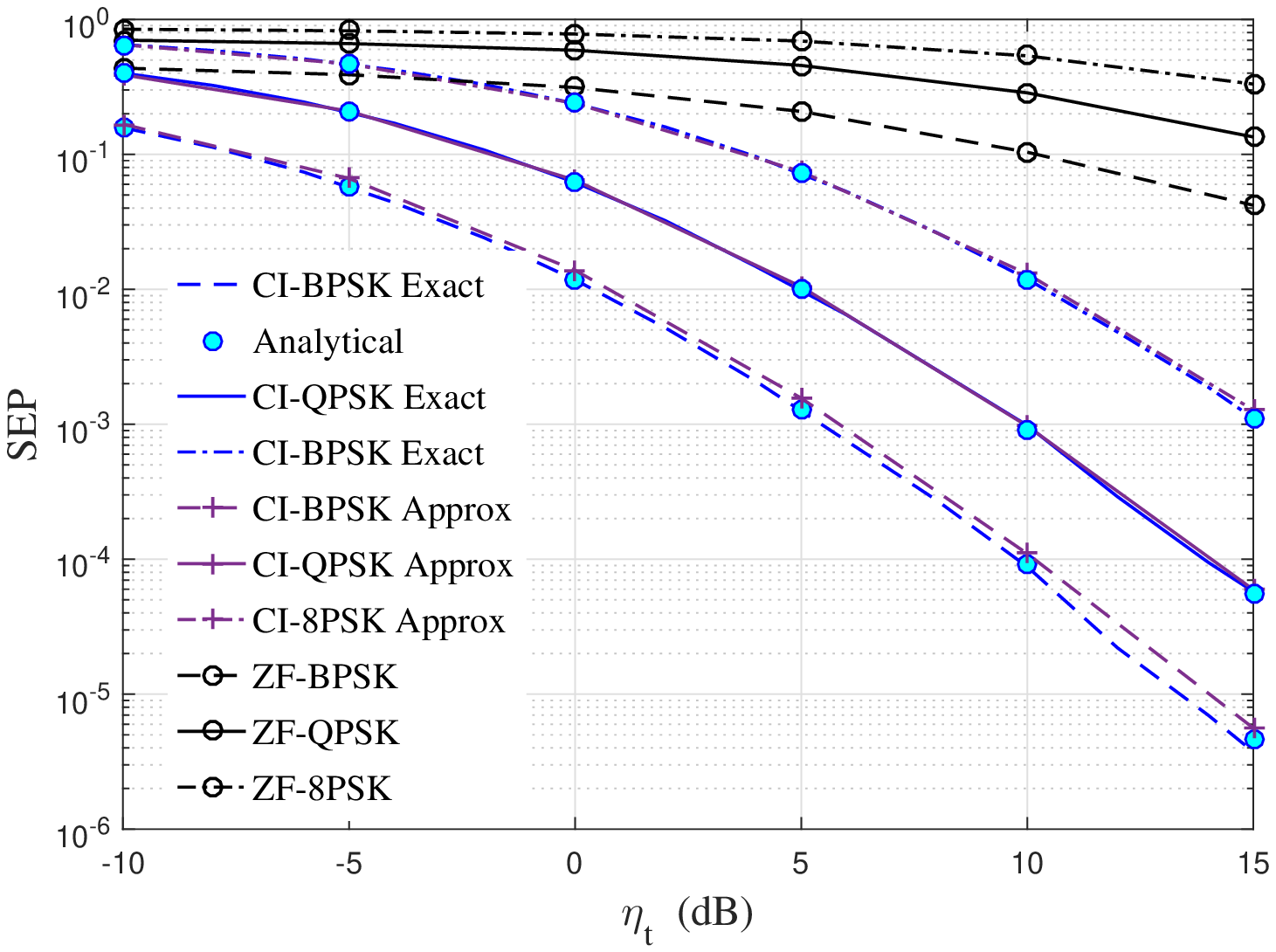}
\par\end{centering}

}\subfloat[\label{fig:5b}SEP versus transmit SNR, $\eta_{t}$, with different
types of input, when $N=8,\, K=6$.]{\noindent \begin{centering}
\includegraphics[scale=0.5]{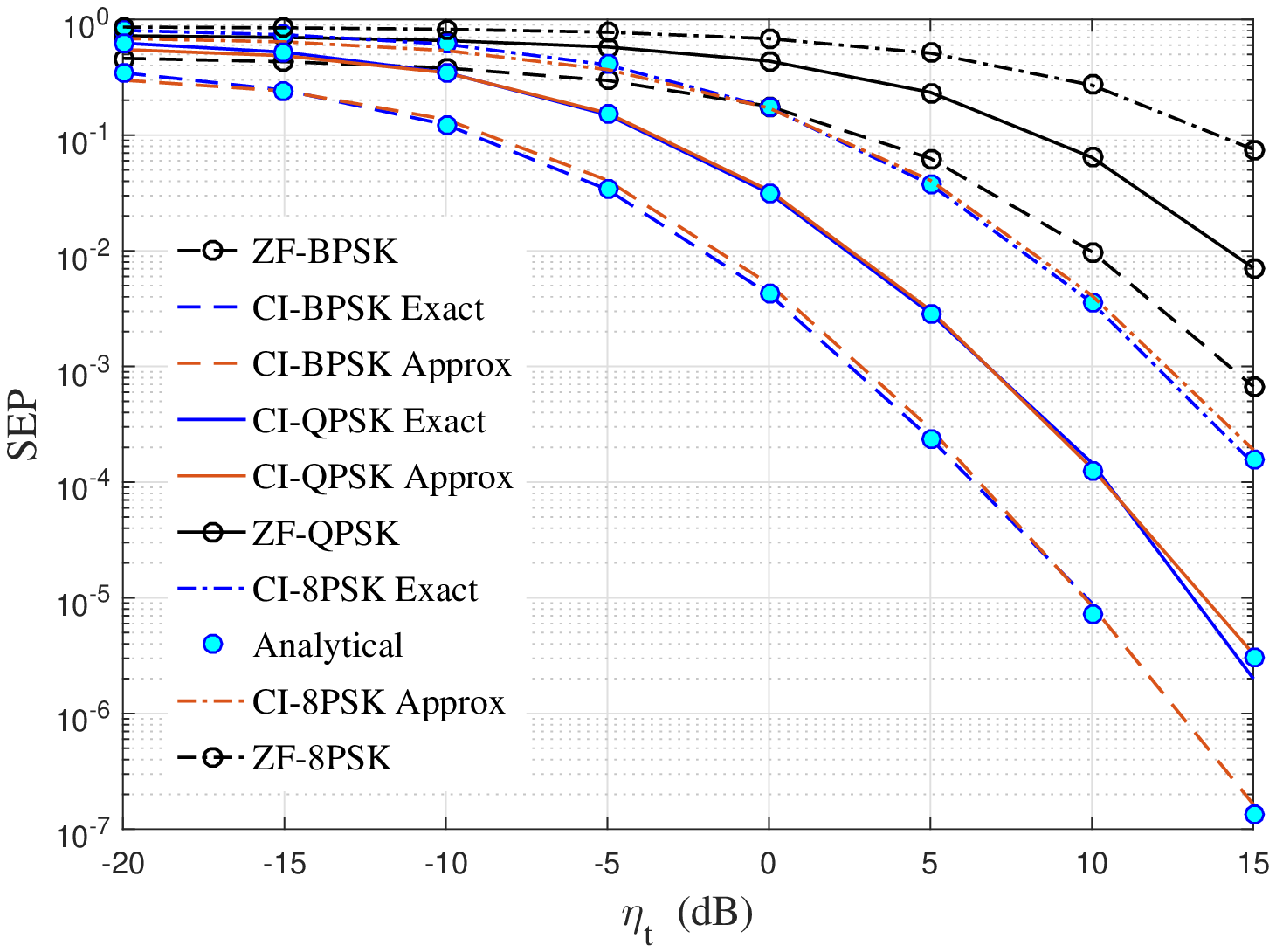}
\par\end{centering}

}
\par\end{centering}

\protect\caption{\label{fig:5}SEP versus transmit SNR for various input types, when
$N=6,8\textrm{ and }K=6$.}
\end{figure}

Fig. \ref{fig:4}, shows the exact and approximated average SEP versus
transmit SNR, $\eta_{t}$, for different types of input, BPSK, QPSK
and 8-PSK. Fig. \ref{fig:4a}, presents the average SEP when $N=K=4$,
and Fig. \ref{fig:4b}, illustrates the average SEP when $N=6,\,\textrm{ and }K=4$.
Additionally and for seek of comparison, some results of the conventional
interference suppression, ZF, technique are also included in these
figures. It should be pointed out that the analytical results in these
figures are obtained from the expressions derived in Section (\ref{sec:Average-Symbol-Error}).
Several interesting points can be extracted from this figure. Firstly,
it is evident that the SEP reduces with increasing the transmit SNR,
$\eta_{t}$, and CI precoding technique always outperforms the ZF
technique in the all SNR values with an up to 15dB gain in the transmit
SNR for a given SEP. In addition, it is clear that the approximated
results obtained from Section (\ref{sub:Approximate-SEP}) are very
tight to the exact ones. Finally, comparing Fig. \ref{fig:4a} and
Fig. \ref{fig:4b}, we can see that, increasing number of BS antennas
always enhances the average SEP, and reduces the gap performance between
the two precoding techniques.

In order to investigate the impact of number of users and number of
BS antennas on the average SEP, in Fig. \ref{fig:5} we present the
average SEP for the CI and ZF precoding techniques for BPSK, QPSK
and 8PSK, when $N=K=6$, as in Fig. \ref{fig:5a} and when $N=8,K=6$
as in Fig. \ref{fig:5b}. From the results in Figs. \ref{fig:5} and
\ref{fig:4}, it is obvious that increasing number of BS antennas
$N$ and/or number of users $K$ lead to enhance the system performance.
Furthermore, the CI precoding has always better performance than ZF
in the all SNR values with an up to 20dB gain in the transmit SNR
for a given SEP. In addition, comparing the average SEP in Fig. \ref{fig:5a}
and Fig. \ref{fig:5b}, similar observations can be concluded as in
the previous case when $K=4$. 

\begin{figure}
\begin{centering}
\subfloat[\label{fig:6a}SEP versus transmit SNR, $\eta_{t}$, with different
power allocation schemes and QPSK input, when $N=K=3$.]{\begin{centering}
\includegraphics[scale=0.5]{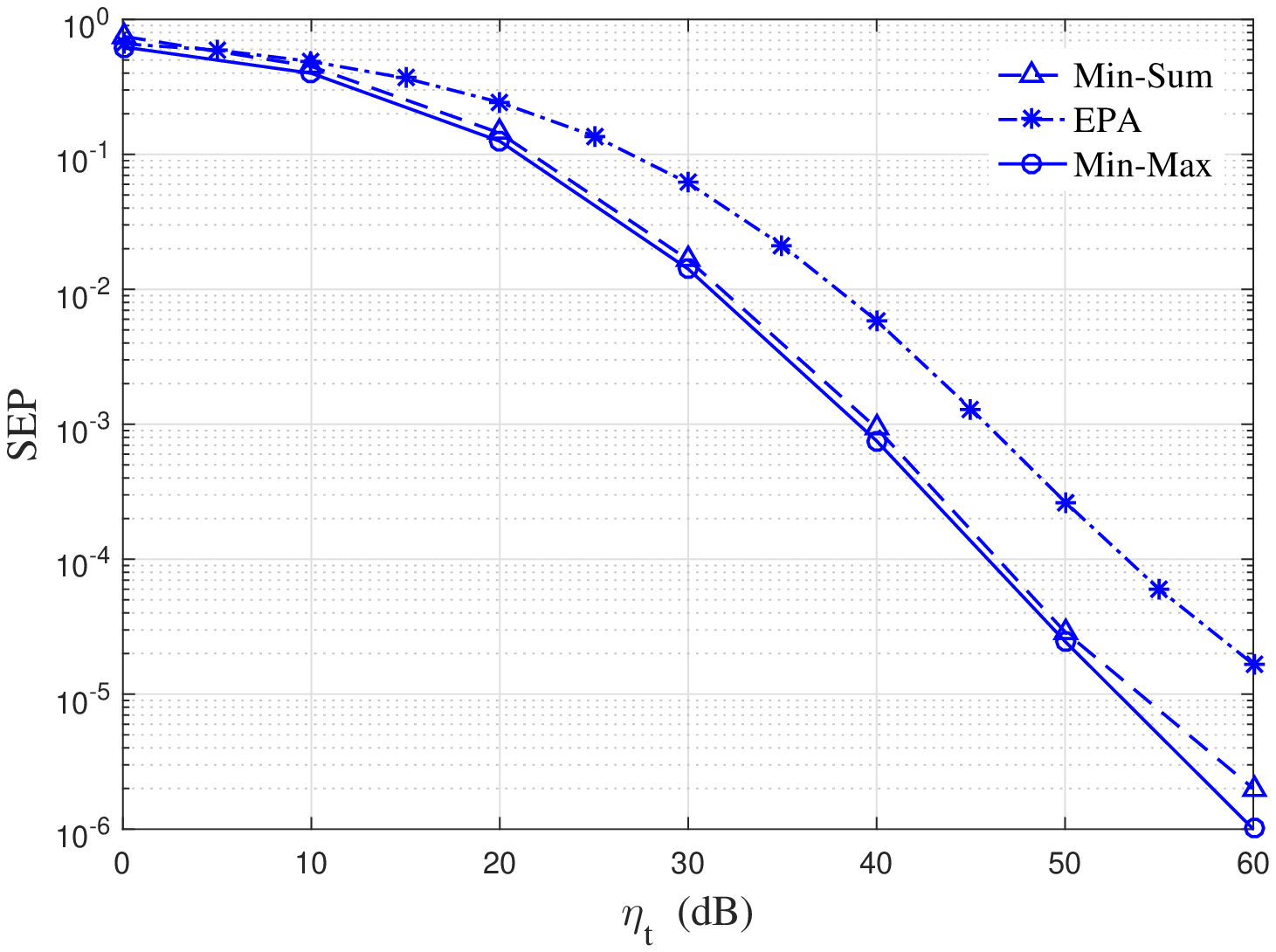}
\par\end{centering}

}\subfloat[\label{fig:6b}SEP versus transmit SNR, $\eta_{t}$, with different
power allocation schemes and QPSK input, when $N=K=8$.]{\begin{centering}
\includegraphics[scale=0.5]{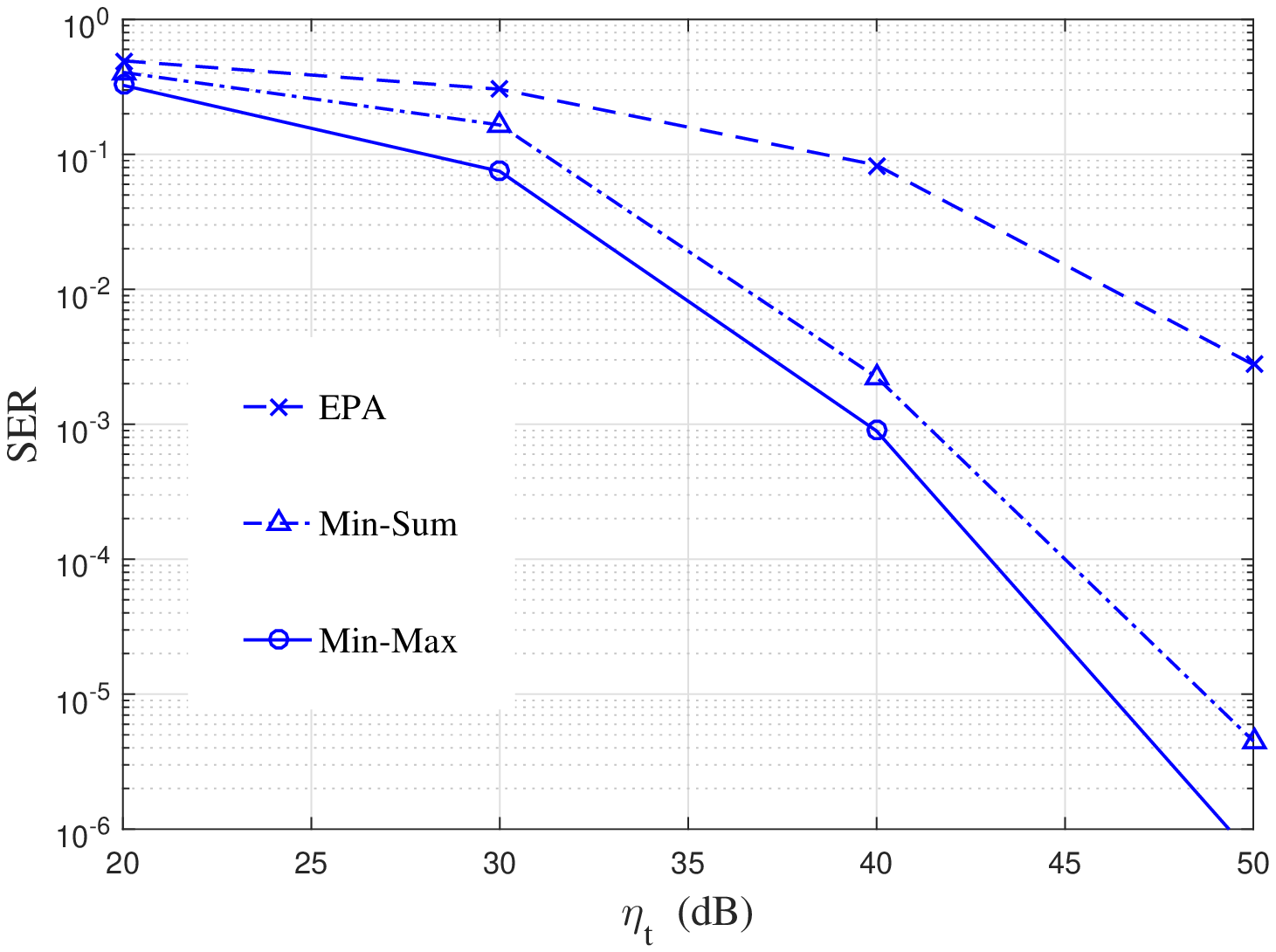}
\par\end{centering}

}
\par\end{centering}

\protect\caption{\label{fig:6}SEP versus transmit SNR with different power allocation
schemes.}
\end{figure}

Fig. \ref{fig:6} illustrates the average SEP versus the transmit
SNR, $\eta_{t}$, for different power allocation schemes, EPA, Min-Sum
and Min-Max schemes. Fig. \ref{fig:6a}, presents the average SEP
versus $\eta_{t}$ when $N=K=3$, while Fig. \ref{fig:6b}, presents
the average SEP versus $\eta_{t}$ when $N=K=8$. From this figure
it can be observed that, EPA scheme always results in the highest
SEP in the all cases. Therefore, we can say EPA scheme provides the
lower bound of the average SEP for the considered MU-MIMO system.
In addition, looking closer at the results in Fig. \ref{fig:6a} and
Fig. \ref{fig:6b} one can clearly observe that, the SEP is dominated
by the performance of the worst user, and thus the Min-Max scheme
has the best performance. It is also noted that, in low transmit SNR
values Min-Sum scheme allocates most the transmission power to the
best/ closest user to the BS and small amount of power to the farther
users, whilst Min-Max scheme allocates relatively high transmission
power to the farther user and small amount of power to the near users.
In addition, as the transmit SNR value increases Min-Sum scheme starts
gradually increasing the power allocated to the farther users at the
expense of the power allocated to the near users.

\begin{figure}
\noindent \begin{centering}
\includegraphics[scale=0.6]{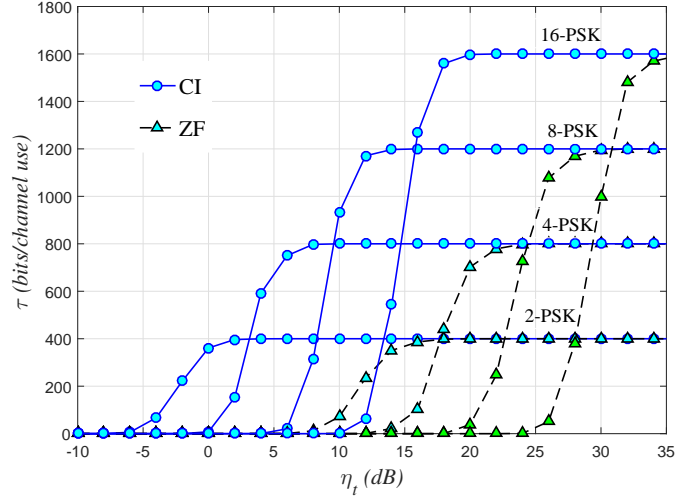}
\par\end{centering}

\protect\caption{\label{fig:6a-1}Throughput versus transmit SNR, $\eta_{t}$, for
various input types, when $\mathscr{N}=100,$ $Q=5$, and $N=K=4$.}
\end{figure}

\noindent In Fig. \ref{fig:6a-1} we present the throughput versus
the transmit SNR, $\eta_{t}$, for different types of input, BPSK,
QPSK, 8-PSK and 16-PSK. For seek of comparison, results of the conventional
ZF precoding technique are included in the figure. The results in
this figure are obtained from the expressions provided in Section
(\ref{sec:Throughput-and-Power}). It is evident that the throughput
saturates to the value of, $\log_{2}\left(M\right)\times\mathscr{N}\times K$,
past a certain transmit SNR $\eta_{t}$ value, the throughput saturates
at 400 bits/channel use in BPSK, at 800 bits/channel use in QPSK,
at 1200 bits/channel use in 8-PSK and at 1600 bits/channel use in
16-PSK. In addition, the CI precoding outperforms the conventional
ZF scheme for a wide range with an up to 15dB gain in the transmit
SNR for a given throughput value. Finally and as anticipated, in low
SNR values the lower modulation orders have better performance than
the higher ones, for instance at 0 dB BPSK achieves the highest throughput.
However, in high SNR values the higher modulation orders achieve better
performance, for instance at 20 dB 16-PSK has optimal performance. 

\begin{figure}
\noindent \begin{centering}
\subfloat[\label{fig:7a}Power Efficiency versus number of BS antennas, $N$,
for various input types, when $\mathscr{N}=100,$ $Q=10$, and $K=4$.]{\begin{centering}
\includegraphics[scale=0.5]{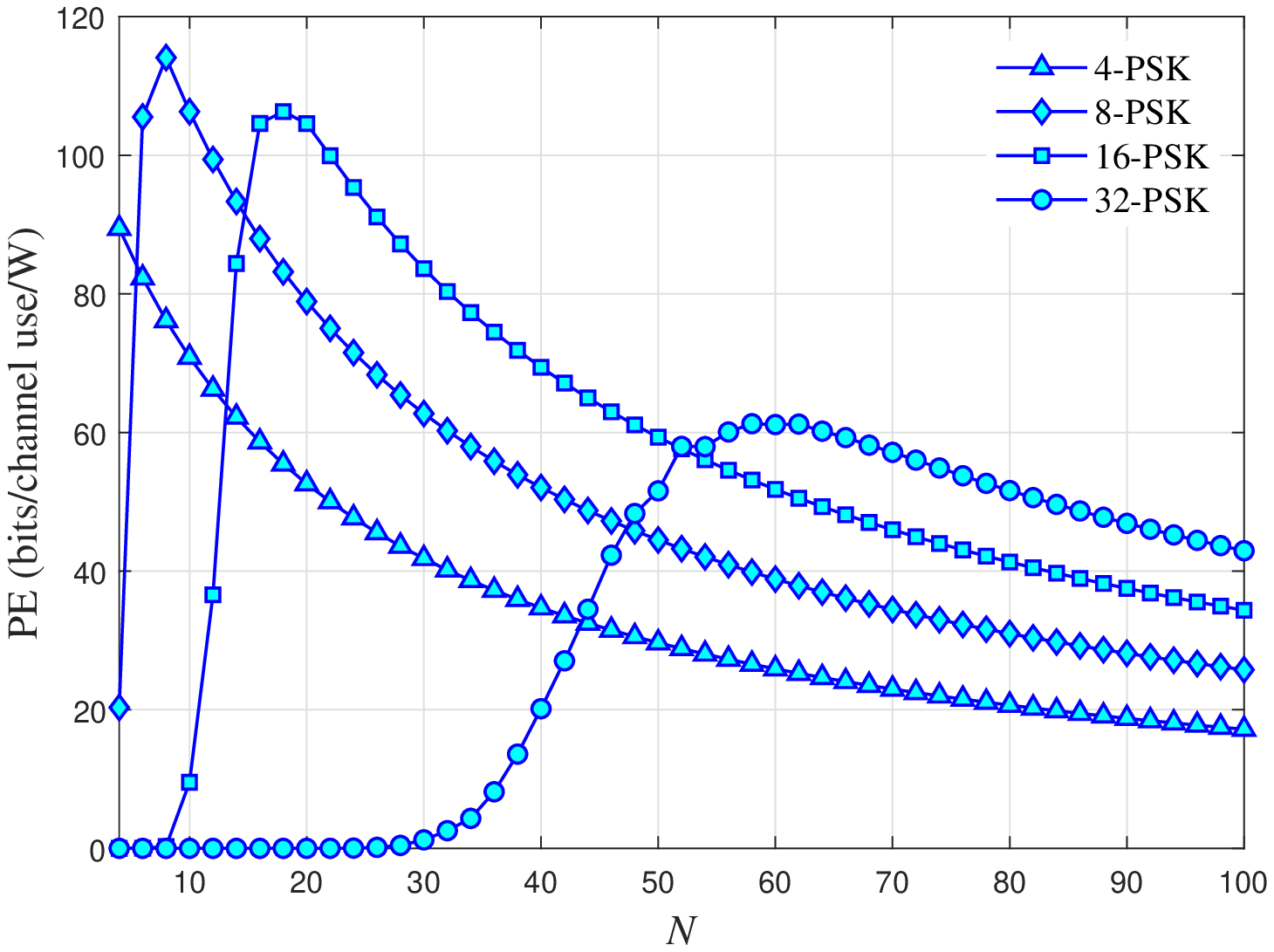}
\par\end{centering}

}\subfloat[\label{fig:7b}Power Efficiency versus number of BS antennas, $N$,
for various input types, when $\mathscr{N}=100,$ $Q=10$, and $K=4$.]{\begin{centering}
\includegraphics[scale=0.5]{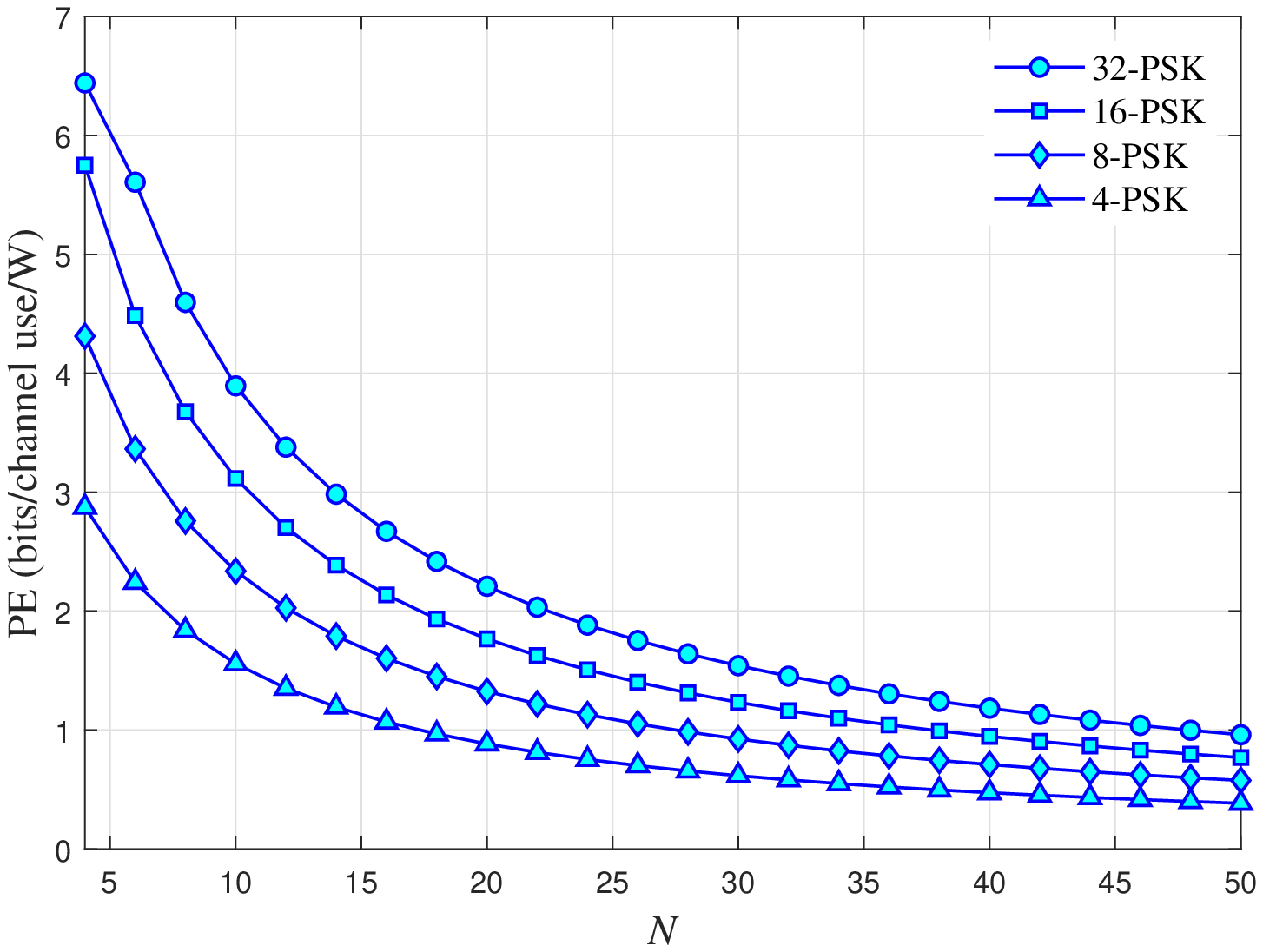}
\par\end{centering}

}
\par\end{centering}

\protect\caption{\label{fig:7}Power Efficiency versus number of BS antennas, $N$,
for different values of the transmission power. }

\end{figure}

\noindent Finally, Fig. \ref{fig:7} depicts the power efficiency
as function of number of BS antennas, $N$, for different values of
the transmission power. The results in these figures are obtained
from the power efficiency expression provided in Section (\ref{sec:Average-Symbol-Error}).
In Fig. \ref{fig:7a} we present the power efficiency versus $N$
when $\varsigma_{DC}$=0.075, $\varsigma_{MS}=0.09$, $P_{P}=35\textrm{ dbm}$,
$\eta_{pa}=0.8$, $P_{D}=7.8\,\textrm{mW}$, $P_{m}=15.2\,\textrm{mW}$,
$P_{f}=10\,\textrm{mW}$, $P_{sy}=25\,\textrm{mW}$, and $P_{DS}=2\textrm{W}$
\cite{powerconsmain,powercons60,powercons62}. From Fig. \ref{fig:7a}
we can observe that when number of BS antennas is small the lower
modulation orders achieve higher power efficiency than the higher
orders, for instance when $N=4$ QPSK has best performance. On the
other hand, when number of BS antennas is large the higher modulation
orders become better than the lower ones, for instance 32-PSK achieves
the highest power efficiency when $N=60$. Furthermore, in order to
clearly demonstrate the impact of transmission power on the power
efficiency for different types of input, we plot in Fig. \ref{fig:7b}
the power efficiency versus $N$ when the transmission power is very
high $P_{P}=20\textrm{ dbW}$. In this case, the higher modulation
orders always have better system performance regardless number of
antennas implemented at the BS. Furthermore, comparing Figs. \ref{fig:7a}
and \ref{fig:7b} it can be concluded that, the power efficiency achieved
in low transmit SNR is much higher than that in high transmit SNR
regime.

\section{Conclusions\label{sec:Conclusions}}

In this paper the statistics of the received SNR of CI precoding technique
has been considered for the first time. Firstly, exact closed form
expressions of the MGF and the average received SNR have been derived.
Then, the derived MGF expression was used to calculate the average
SEP. In light of this, exact average SEP expression for CI precoding
with $M$-PSK was obtained. In addition, accurate asymptotic approximation
for the average SEP has been provided. Building on the new performance
analysis, different power allocation schemes to enhance the average
SEP have been considered. In the first scheme, power allocation technique
based on minimizing the total SEP was studied, while in the second
scheme power allocation technique based on minimizing the maximum
SEP was investigated. Furthermore, new and explicit analytical expressions
of the throughput and power efficiency of the CI precoding in MU-MIMO
systems have been derived. The results in this paper explained that
the CI scheme outperforms ZF scheme in the all considered metrics.
Furthermore, increasing the transmit SNR, number of users and number
of BS antennas always enhance the achieved SEP. It was also shown
that, using EPA leads to the highest SEP and the considered power
allocation techniques can perform very low SEP. Finally, in low transmit
SNR values and when number of BS antennas is small, the lower modulation
orders achieve higher power efficiency than the higher modulation
orders. 

\bibliographystyle{IEEEtran}
\bibliography{bib}

\end{document}